\documentclass[aps,superscriptaddress,rmp,floatfix,twocolumn,10pt]{revtex4-1}
\usepackage{graphicx}
\usepackage{dcolumn}
\usepackage{amsmath}
\usepackage{amsfonts}
\usepackage{amssymb}
\usepackage{epsfig,float,afterpage,wrapfig,psfrag}
\usepackage{color}

\newcommand{\beq}{\begin{equation}}
\newcommand{\eeq}{\end{equation}}
\newcommand{\bes}{\begin{subequations}}
\newcommand{\ees}{\end{subequations}}
\newcommand{\bea}{\begin{eqnarray}}
\newcommand{\eea}{\end{eqnarray}}
\newcommand{\ba}{\begin{array}}
\newcommand{\ea}{\end{array}}
\newcommand{\beqn}{\begin{eqnarray*}}
\newcommand{\eeqn}{\end{eqnarray*}}

\newcommand{\f}[2]{\frac{#1}{#2}}
\newcommand{\g}{\gamma}

\newcommand{\om}{\omega}

\newcommand{\la}{\langle}
\newcommand{\dg}{\dagger}
\newcommand{\ra}{\rangle}

\newcommand{\vg}{v_g}
\newcommand{\veit}{v_\text{EIT}}
\newcommand{\ODB}{\text{OD}_\text{B}}
\newcommand{\OD}{\text{OD}}
\newcommand{\rb}{r_\text{B}}

\def\b0{{\bf 0}}

\def\nn{\nonumber}

\begin{document}
\title{Strongly interacting photons in one-dimensional continuum}

\author{Dibyendu Roy}
\affiliation{Max Planck Institute for the Physics of Complex Systems, N{\"o}thnitzer Str. 38, 01187 Dresden, Germany}
\affiliation{Raman Research Institute, Bangalore 560080, India}
\altaffiliation
{Present address}

\author{Christopher M. Wilson}
\affiliation{Electrical \& Computer Engineering Department and the Institute for Quantum Computing, University of Waterloo, ON N2L 3G1, Canada}

\author{Ofer Firstenberg}
\affiliation{Department of Physics of Complex Systems, Weizmann Institute of Science, Rehovot 76100, Israel}

\begin{abstract}
Photon-photon scattering in vacuum is extremely weak. However, strong effective interactions between single photons can be realized by employing strong light-matter coupling. These interactions are a fundamental building block for quantum optics, bringing many-body physics to the photonic world and providing important resources for quantum photonic devices and for optical metrology. In this Colloquium, we review the physics of strongly-interacting photons in one-dimensional systems with no optical confinement along the propagation direction. We focus on two recently-demonstrated experimental realizations: superconducting qubits coupled to open transmission lines, and interacting Rydberg atoms in a cold gas. Advancements in the theoretical understanding of these systems are presented in complementary formalisms and compared to experimental results. The experimental achievements are summarized alongside a description of the quantum optical effects and quantum devices emerging from them.
\end{abstract}

\vspace{0.0cm}
\maketitle
\tableofcontents
\section{Introduction}
\label{intro}
Photons, the carriers of the electromagnetic force, are elementary particles
with no charge and zero rest mass. Photon-photon scattering in vacuum is extremely weak \cite{Schwinger51, Karplus51, Iacopini79} and has, in fact, never been experimentally observed at optical or lower frequencies \cite{Zavattini2012}.  This makes photons  excellent long-distance carriers of classical and quantum information. However, this seclusion of photons poses substantial challenges for efficiently employing them for information processing. In classical communication networks, optical signals are often converted to electrical signals, which can then be manipulated using solid-state devices.  However, existing conversion methods are inefficient at the few-photon level and are ill-suited for quantum information processing. This has motivated scientists in a number of fields, such as \emph{nonlinear quantum optics} and \emph{cavity quantum electrodynamics}, to study effective photon-photon interactions \cite{Gibbs85} with the ultimate goal of strong and controllable coupling between single photons. Interactions at the single-photon level are essential for a wide variety of quantum-optical applications. For instance, they form the basis for all-optical quantum gates \cite{ImamogluPRL1997} and enable metrology beyond the standard quantum limit \cite{Mitchell11}. They are also important from the fundamental physics viewpoint, endowing photonic systems with matter-like properties and realizing strong correlations and quantum many-body behavior in light \cite{Carusotto13,Chang08}.

\begin{figure}[tb]
\includegraphics[trim={1cm 0 0 4cm},clip,width=8.5cm]{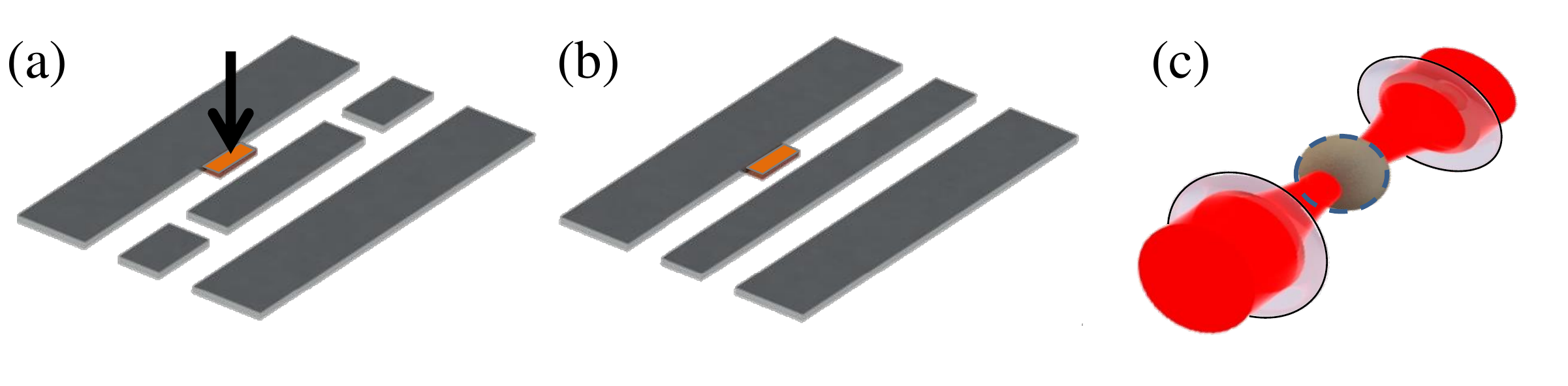}
\caption{(Color online). (a) A superconducting qubit (marked with an arrow) embedded in a one-dimensional transmission line waveguide. A cavity is formed by two capacitive gaps in the middle conductor. (b) Without the cavity. (c) Probe light focused through a dense atomic cloud, exciting a single Rydberg atom within the blockade sphere (dashed line).}
\label{cqed}
\end{figure}

Cavity quantum electrodynamics (QED) is a paradigmatic discipline
\cite{Walther06, Miller05, Haroche13} exhibiting effective photon-photon
interactions in the quantum regime \cite{Birnbaum05, Schuster08}. In cavity
QED, atoms are placed inside a high-finesse electromagnetic resonator, in
which the radiation spectrum is discrete. Bouncing between the resonator mirrors, a single photon interacts with the atoms
effectively many times, significantly enhancing the atom-photon coupling. This in turn can
generate strong correlations between the photons. Cavity QED experiments with
free atoms have been carried out with alkali metals in optical cavities
\cite{Mabuchi02,Thompson92,Birnbaum05,Schuster08,Dayan08} and with Rydberg
atoms in microwave cavities \cite{Nogues99, Guerlin07, Deleglise08,
  Raimond01}. Cavity QED experiments have also been conducted on a variety of
solid-state systems, including quantum dots (QDs) in photonic crystals
\cite{Yoshie04, Badolato05, Hennessy07, Englund07, Fushman08,
  Carter2013} and superconducting microwave circuits.  In the latter, known as
circuit QED \cite{Girvin09,Wallraff04,Chiorescu04, Blais04}, superconducting qubits acting as artificial atoms are coupled to microwave photons in Fabry-P\'{e}rot cavities made of coplanar waveguides as shown in Fig.~\ref{cqed}(a). Both the fields of cavity and circuit QED have been very successful, providing both significant fundamental results and important advances in quantum information science. Nevertheless, the cavities used to enhance the coupling in these systems also present several disadvantages, for instance, the narrow
bandwidth of the emitted photons and the problem of stochastic release of photons by the cavity \cite{Hoi12}.  A related challenge is the coupling of photons into and out of the cavities with high efficiency, as is required to link large numbers of nodes in quantum networks \cite{Aoki09}.

Because of these limitations, much recent work has focused on cavity-free systems. The coupling strength in these system can be quantified by the extinction of a propagating photon by a single emitter $(1-T)$, where $T$ is the transmission coefficient. Although the nomenclature is still settling, a reasonable definition of strong coupling in these open systems is that $1-T>50\%$, which implies that the emission rate from the
atom into the desired  mode is larger than the decoherence rate associated with
all other process, including emission into other modes.
\textcite{vanEnk01} and \textcite{Zumofen08} showed theoretically that a single atom can fully block ($1-T=100\%$) photons in open space, if their spatial and temporal mode matches the atomic radiation pattern, while a tightly focused beam is limited to $1-T<85\%$. In both cases, photons are transversely focused to an area $A$ comparable with the scattering cross-section of the atom $\sigma_a$, and their electric field becomes large enough to excite the atom with near unity probability. 
The highest extinction by a single emitter experimentally achieved in three-dimensional (3D) open space is $1-T=30\%$ \cite{maser2016few}.

A major barrier to higher extinction in open space is the spatial-mode mismatch between the incident and scattered waves. This problem has recently been solved in two complementary ways, leading to strong coupling  and photon-photon interactions in cavity-free one-dimensional (1D) systems:

(i) \emph{Superconducting qubits in open transmission lines}, which are
related to circuit QED systems, as illustrated in Fig.~\ref{cqed}(b). These
systems enhance the coupling in two ways, both inherited from circuit QED.
Most importantly, superconductors can confine the microwave fields to deeply
subwavelength sizes in the transverse dimensions. This produces a mode volume,
in units of cubic wavelengths, that is orders of magnitude smaller than that
of 3D cavities or free fields. In addition, transition dipoles of
superconducting qubits are much larger than those of real atoms. These effects
together allow for the observation of strong coupling \cite{Astafiev10a}, extinctions of $1-T>99\%$ \cite{Hoi11}, and strong photon-photon correlations \cite{Hoi12}.

(ii) \emph{Rydberg atoms excited by focused optical beams in a dense atomic gas.} In these systems, strong coupling is achieved by greatly enhancing the size of the effective scatterer, thereby achieving mode matching to collimated light beams. The dipolar interaction between Rydberg atoms prevents the excitation of more than one Rydberg atom inside the volume of a so-called \emph{blockade} sphere. With only zero or one excitation, each blockade sphere thus acts as a ``superatom" \cite{Vuletic06}. The weak coupling of photons to each individual atom can sum up in a dense gas to a strong effective coupling with the superatoms, leading to extinction of $1-T\ge95 \%$ \cite{DurrPRL2014}.
As long as the blockade sphere is wider than the beam waist, as illustrated in Fig.~\ref{cqed}(c), the evolution is limited to the longitudinal 1D continuum. Photon-photon interactions were observed in this system, with photons either blocking \cite{Peyronel12} or spatially attracting \cite{Firstenberg13} each other.

These cavity-free systems feature intrinsically-nonequilibrium, quantum many-body dynamics. The input field is driven by either a laser or microwave generator, imposing a nonequilibrium boundary condition on the propagating photons in 1D. Therefore, the study of photon-photon correlation mediated by local light-matter coupling in 1D calls for advanced quantum field theories in the strongly-interacting and nonequilibrium regimes. Traditionally, photon transport in this type of
system is studied by employing a master equation that assumes a weak coherent
state as input and usually involves approximations such as linearization of operator equations (see Sec.~\ref{sec-inout} for details) and the Markovian approximation \cite{Agarwal13}. Here, we review the recent progress in
developing new analytical and numerical techniques to study collective
scattering of multiple photons from two-, three-, or multi-level emitters in a
1D continuum.

This Colloquium presents an overview of this research, emphasizing the systems discussed above. Other 1D systems with artificial atoms, which are outside the scope of the paper,
include QDs coupled to surface plasmons of a metallic
nanowire \cite{Akimov07, Akselrod14, Versteegh14} or to line-defects in
photonic crystals \cite{Laucht12, Lodahl13, Arcari14, Javadi15}, and QDs or
nanocrystals coupled to semiconductor or diamond nanowires \cite{Claudon10, Babinec10,Reithmaier14}. In addition, strong coupling to single emitters in 1D can also be achieved in an ion trap \cite{Meir14}, with cold atoms trapped inside \cite{Bajcsy09} or near \cite{Vetsch10} an optical fiber, or with single molecules doped in an organic crystal inside a glass capillary \cite{Faez14}.

In Sec.~\ref{SinEm}, we summarize the theoretical approaches and experimental results for
systems with single emitters, along with a systematic description of various phenomena and their application to quantum information processing.
Theories and experiments with multiple emitters are presented in Sec.~\ref{MultEm} in the strong-coupling regime, and in Sec.~\ref{sec-Rydberg} in the weak-coupling regime in systems of interacting Rydberg atoms. We conclude with a short
discussion on current research challenges in Sec.~\ref{Conc}.

\section{Single emitter}
\label{SinEm}

\begin{figure}[tb]
\includegraphics[width=8.5cm]{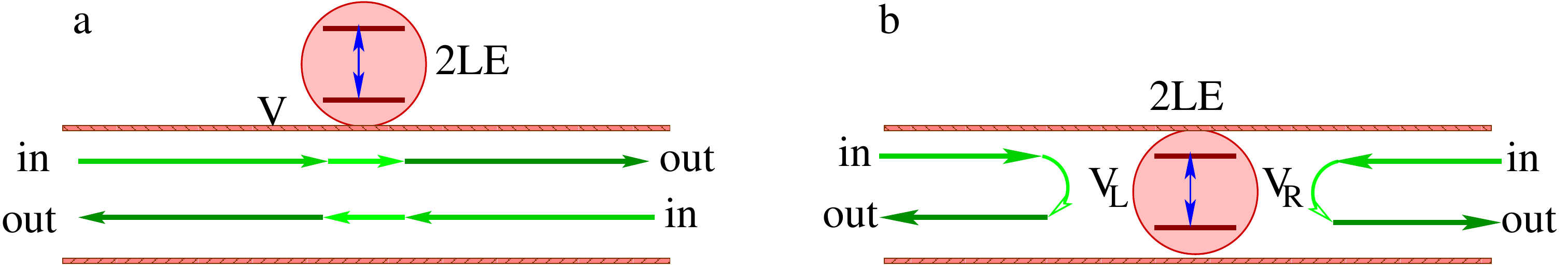}
\caption{(Color online). Two configurations of emitter-photon coupling in an open waveguide. (a) Side-coupled, (b) direct-coupled.}
\label{conf}
\end{figure}

A model configuration of a two-level emitter (2LE) side coupled to photons in
a waveguide is shown in Fig.~\ref{conf}(a). This is a common model for
superconducting qubits coupled to a transmission line \cite{Wallraff04,
  Astafiev10a,Hoi12} and for QDs coupled to surface plasmons
\cite{Akimov07}. The model in Fig.~\ref{conf}(b) has the 2LE directly coupled
to the photons in the waveguide. This model is popular for atomic cavity QED
experiments \cite{Birnbaum05} and is also often used in experiments with
line-defect photonic crystals \cite{faraon07,Park08}. The transmission and reflection of photons
in the side-coupled configuration can be mapped to those in the directly coupled configuration \cite{Shen09}.

Our model system also connects to a number of problems important in condensed matter physics, for instance, the general problem of quantum tunnel in open (dissipative) systems \cite{Caldeira1983, Costi1999} and specifically the spin-boson problem \cite{Leggett1987}, which have seen renewed importance in describing decoherence in various implementations of quantum bits. In the systems of interest in this colloquium, the photons in the 1D continuum play the role of the bosonic bath in the condensed-matter models. There are also exact analogies to various other nonequilibrium quantum impurity models, including the nonequilibrium Kondo model \cite{Gordon98, Cronenwett98, Meir91, Mehta06, Dhar08, Nishino11}.  We note, however, that in the condensed matter context, the models are most often concerned with the dynamics of the spin or impurity. In the present work, we are most often concerned with the dynamics of the bath itself, that is, the photons in the 1D continuum. In addition, the bosonic fields considered in most condensed-matter models are actually collective excitations of matter, such as phonons.  In that sense, experiments related to these models are not generally concerned with light-matter coupling.

A general Hamiltonian of a 2LE side-coupled to photons in a 1D continuum is given by $\tilde{H}=\tilde{H}_0+\tilde{H}_1$, where
\bea
\tilde{H}_0&=&\int dk~\hbar\omega_k a^{\dg}_ka_k+\hbar(\tilde{\omega}_e-i\gamma)|e\ra \la e|,\label{sc2LE}\\
\tilde{H}_1&=&\int dk~\hbar V_{k}(a^{\dg}_k|g\ra \la e|+|e\ra \la g|a_{k}).\label{sc2LEb}
\eea
The first term in Eq.~(\ref{sc2LE}) represents the propagating photon fields of
frequency $\om_k$ and wavevector $k$. The 2LE is described by the second term
in Eq.~(\ref{sc2LE}), with transition frequency $\tilde{\omega}_e$ between states $|g\ra$ and $|e\ra$. The $i\gamma$ term accounts for spontaneous emission into photon modes outside of the 1D continuum, which dominates in atomic systems~\cite{Shen09}. In Subsections \ref{SinEm}B and \ref{SinEm}E, we discuss how to treat pure dephasing, which dominates in superconducting systems. The interaction
of the propagating photons with the 2LE is governed by $\tilde{H}_1$, which is written in the rotating-wave approximation. This is valid for typical light-matter coupling strengths available in recent
experiments. Here $a_{k}~(a^{\dg}_{k})$ is the photon annihilation (creation)
operator, and the coupling strength of a photon of wavevector
$k$ with the emitter is $V_k$. 

\begin{figure}[tb]
\includegraphics[width=5.0cm]{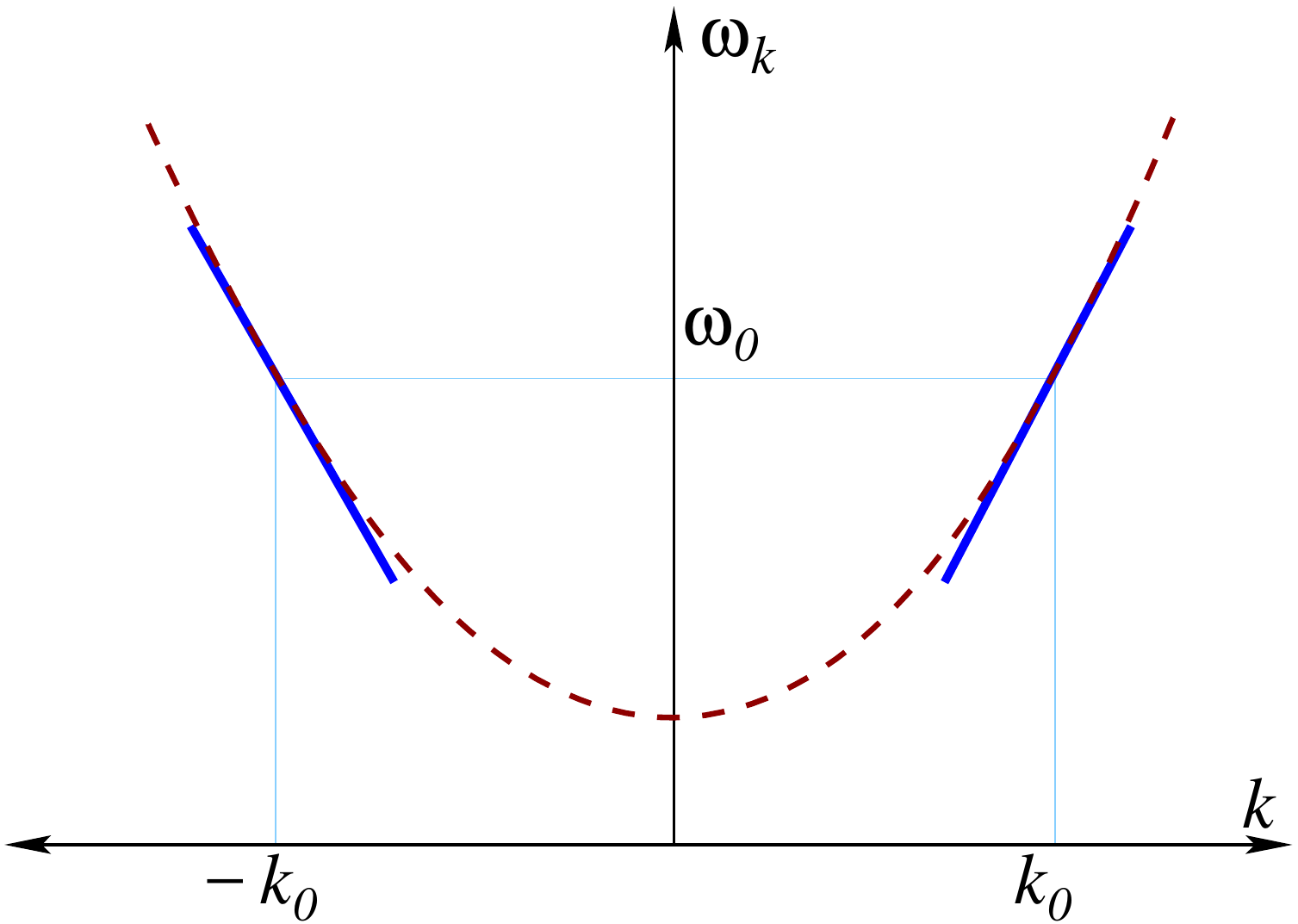}
\caption{(Color online). Linearization of the dispersion relation of the waveguide mode. The full dispersion relation $\om_k$ is shown by the dashed curve. The linearized relations are denoted by the two solid lines around the wavevectors $\pm k_0$, corresponding to a frequency $\om_0$.}
\label{LinDis}
\end{figure}

The energy-momentum dispersion ($\om_k$ versus $k$) of photons in various 1D
 waveguides is generally nonlinear and depends on the properties of
 the waveguide. However, it is convenient to assume linear
 dispersion to describe the first two theoretical approaches
 discussed here.  We can linearize the dispersion near some arbitrary
 frequency $\om_0$ with the
 corresponding wavevector $\pm k_0$ as shown in Fig.~\ref{LinDis}. The approximate linearized dispersion of $\om_k$ around $k_0$ (right-moving photons) and $-k_0$ (left-moving photons) reads
\bea
&&\int_{k\simeq k_0}\om_{k}a^{\dg}_ka_k \simeq \int_{k\simeq k_0}(\om_0+\vg(k-k_0))a^{\dg}_{R,k}a_{R,k}, \label{rmph}\\
&&\int_{k\simeq -k_0}\om_{k}a^{\dg}_ka_k \simeq \int_{k\simeq -k_0}(\om_0-\vg(k+k_0))a^{\dg}_{L,k}a_{L,k}, \nn
\eea
where $a^{\dg}_{R,k}$ ($a^{\dg}_{L,k}$) creates a right-(left-)moving photon
and $\vg$ is the group velocity of photons at $\om_0$. Thus we divide
the propagating photons into two oppositely moving modes (channels). Next, we extend the limits of the integration over $k$ to $(-\infty,\infty)$ for the left and right-moving photons, as we are only interested in photons with a narrow bandwidth in the vicinity of $\om_0$, and we make a change of variables $k \mp k_0 \to k$ for the right- and left-moving photons.

The total excitation operator $N_E=\int dk~[a^{\dg}_{R,k}a_{R,k}+a^{\dg}_{L,k}a_{L,k}]+|e\ra \la e|$
commutes with the linearized Hamiltonian $\tilde{H}$. Subtracting the term $\hbar\om_0N_E$ from the linearized $\tilde{H}$ gives the final Hamiltonian $H=\tilde{H}-\hbar\om_0N_E=H_0+H_1$, where
\bea
\frac{H_0}{\hbar}&=&\int_{-\infty}^{\infty} dk~
\vg k(a^{\dg}_{R,k}a_{R,k}-a^{\dg}_{L,k}a_{L,k})+(\om_e-i\gamma)|e\ra \la e|,\nn\\
\frac{H_1}{\hbar}&=&\int_{-\infty}^{\infty} dk[V_k(a^{\dg}_{R,k}+a^{\dg}_{L,k})|g\ra \la e|+h.c.],\label{HamL}
\eea
with $\om_e=\tilde{\om}_e-\om_0$.

Several interrelated theoretical techniques have been employed in recent years for investigating correlated photon dynamics in a 1D continuum. We can
divide them into five different groups: (a) multiparticle scattering theory in a continuum \cite{Shen07a, Shen07b, Yudson08, Roy10a, Zheng10}, (b) the input-output
formalism of quantum optics \cite{Fan10, Koshino12, Peropadre13}, (c)
an approach based on the Lippmann-Schwinger equation \cite{Roy11b, Zheng13a}, (d) a method based on the
Lehmann-Symanzik-Zimmermann reduction for the multiphoton scattering process
\cite{Shi09}, and (e) the time-dependent, wave-packet evolution approach
\cite{Longo10}. Scattering theory is a well-known framework to study scattering of waves and particles within the Schr{\"o}dinger picture of quantum mechanics, and it has been extensively applied in different branches of physics. The input-output formalism was mainly developed for understanding light-matter interaction, and it is based on the Heisenberg picture.  In what follows, we carefully discuss these two
techniques and the connection between them. We briefly mention the applications of the other approaches later on.

\subsection{Scattering theory}
\label{scatt}
In scattering theory, the scattering matrix $S$ expresses how an incoming state of
monochromatic, free photons evolves via a local interaction with atoms into a
superposition of outgoing monochromatic photons. The $S$-matrix for scattering
of an $N$-photon state is defined by \cite{Taylor06, Newton82}
\bea
S^{(N)}_{{\bf p};{\bf k}}=\la {\bf p}|S|{\bf k}\ra,
\eea
where $|{\bf k}\ra$ and $|{\bf p}\ra$ are incoming and outgoing photonic
states respectively. Here, the vectors ${\bf k}$ and ${\bf p}$ denote the
incoming and outgoing momenta of the $N$ photons. These incoming and
outgoing states are considered to be free states in the interaction picture, and they exist long before ($t_0 \to -\infty$) and long after ($t_1 \to \infty$)
the scattering occurs. The operator
\bea
S&=&\lim_{\substack{t_0 \to -\infty \\ t_1 \to \infty}}U_I(t_1,t_0) \nn
\eea
is given by the time-evolution operator $U_I$ in the interaction picture $U_I(t_1,t_0)=e^{iH_0t_1/\hbar}e^{-iH(t_1-t_0)/\hbar}e^{-iH_0t_0/\hbar},$ and the above S-matrix can be redefined as
\bea
\la {\bf p}|S|{\bf k}\ra= \la {\bf p}^-|{\bf k}^+\ra.\label{smatrix}
\eea
The scattering eigenstates $|{\bf k}^{+}\ra$ and $|{\bf p}^{-}\ra $ evolve in the interaction
picture from a free-photon state either in the distant past or the distant
future :
\bea
|{\bf k}^+\ra=U_I(0,t_0) |{\bf k}\ra=e^{iHt_0/\hbar}e^{-iH_0t_0/\hbar}|{\bf k}\ra=\Omega_+|{\bf k}\ra,
\nn\\
|{\bf p}^-\ra=U_I(0,t_1) |{\bf p}\ra=e^{iHt_1/\hbar}e^{-iH_0t_1/\hbar}|{\bf p}\ra=\Omega_-|{\bf p}\ra,
\nn
\eea
where we drop the limits of $t_0,t_1$ for compactness and imply $t_0 \to
-\infty$ and $t_1 \to \infty$ in all forthcoming similar expressions. It is also
possible to introduce input and output operators $a^{\dg}_{i_m,\rm
  in}(k_m)$ and $a^{\dg}_{o_m,\rm  out}(p_m)$ respectively, which create the incoming and outgoing scattering eigenstates such that
\bea
&&S^{(N)}_{{\bf p};{\bf k}}=\la {\bf p}|S|{\bf k}\ra =\la {\bf p}^-|{\bf k}^+\ra\label{sminout}\\
&&=\la \varphi| a_{o_1,\rm out}(p_1)..a_{o_N,\rm out}(p_N)a^{\dg}_{i_1,\rm
  in}(k_1)..a^{\dg}_{i_N,\rm in}(k_N)|\varphi\ra, \nn
\eea
where $|\varphi \ra$ is the vacuum state, and
\bea
a_{i_m,\rm in}(k_m)&=&\Omega_+a_{i_m,k_m}\Omega_+^{\dg}, \label{stinput}\\
a_{o_m,\rm out}(k_m)&=&\Omega_-a_{o_m,k_m}\Omega_-^{\dg},\label{stoutput}
\eea
with the commutation relations
\bea
&&[a_{i_m,\rm in}(k_m),a^{\dg}_{i_n,\rm in}(k_n)]=\delta(k_m-k_n)\delta_{i_m,i_n}, \nn\\
&&[a_{o_m,\rm out}(p_m),a^{\dg}_{o_n,\rm out}(p_n)]\nn=\delta(p_m-p_n)\delta_{o_m,o_n}.\nn
\eea
The indices $i_m,o_m$ can take on the values $L,R$ for $m=1,2..N$, depending on whether the $m$-th incoming or outgoing photon is left-moving or right-moving.  The connection of these input and output operators to those in the
input-output formalism will become clear in the next subsection.

\textcite{Shen07a, Shen07b} have recently developed a method inspired by the nonperturbative Bethe-ansatz calculation to derive exact scattering eigenstates $|{\bf k}^+\ra$ of a few photons. 
The incoming and outgoing photon states can be obtained from the
scattering eigenstate $|{\bf k}^+\ra$. The dynamics of two photons in this system are very different from those of
a single photon, as they become correlated via the collective scattering from the 2LE \cite{Rupasov84a, Rupasov84b, Deutsch92,Cheng95}.  The approach we present treats the atom-photon dynamics in real space, which is particularly convenient for discussing steady-state photon transport from one space-time point to another.

 We start with an ansatz for the full scattering eigenstate $|{\bf k}^{+}\ra$
 of $H$ for a particular unscattered state $|{\bf k}\ra$ of $H_0$. The total
 number of photons $N$ is conserved during the scattering process when using the rotating-wave approximation. To calculate different
 amplitudes of the scattering eigenstate, we employ the time-independent
 Schr{\"o}dinger equation $H|{\bf k}^{+}\ra=\hbar \vg(k_1+k_2+..+k_N)|{\bf
   k}^{+}\ra$ with boundary conditions that determine the propagation direction of the incident photons. We now write down an
 effective representation of $H$ in real space, where the evolution of the
 incident photons is more conveniently described. To this end, we take the photon operators in momentum space to be the Fourier transforms of real-space operators, for example
\bea
&&a_{R,k}= \f{1}{\sqrt{2\pi}}\int_{-\infty}^{\infty}dx~ a_R(x)e^{-ikx}, \nn
\eea
where $a_R(x)$ annihilates a right-moving photon at position $x$ \cite{Shen09}. Thus, we find an effective real-space Hamiltonian for a 2LE coupled to 1D continuum with linear dispersion,
\bea
&&\frac{H_{\rm eff}}{\hbar}=-i\int dx ~\vg\Big[a^{\dg}_R(x)\f{\partial}{\partial x}a_R(x)-a^{\dg}_L(x)\f{\partial}{\partial x}a_L(x)\Big]\nn\\
&&+(\omega_e-i\gamma)|e\ra \la e|+V\big[(a^{\dg}_R(0)+a^{\dg}_L(0))|g\ra \la e|+{\rm h.c.}\big].\label{sc2LE1}
\eea
We assumed here that the coupling $V_k\equiv V/\sqrt{2\pi}$ is independent of the wavevector $k$ (the Markov approximation). The Hamiltonian
$H_{\rm eff}$ is nonhermitian in the presence of the dissipation term $\gamma$. In the following, we
calculate $|{\bf k}^{+}\ra$ using the hermitian $H_{\rm eff}$ without $\gamma$, and
subsequently replace $\om_e$ by $\om_e-i\gamma$ in the final results \cite{Rephaeli13}.

The incident photons can be injected in the right-moving and/or left-moving channels. The nonequilibrium
dynamics can be probed in experiments by measuring the transmission and reflection of photons at the opposite sides of the waveguide. For a side-coupled 2LE, the transmission coefficient is calculated from the number of photons remaining in the incident channel  (or channels) after scattering, and the reflection coefficient is determined by counting photons in the opposite channel (or channels) after scattering. For example, the transmission and reflection coefficient for $N$ right-moving incident photons are respectively
\bea
T=\f{\la {\bf k}^{+}|a^{\dg}_R(x)a_R(x)|{\bf k}^{+}\ra}{\la {\bf k}|a^{\dg}_R(x)a_R(x)|{\bf k}\ra},~R=\f{\la {\bf k}^{+}|a^{\dg}_L(x')a_L(x')|{\bf k}^{+}\ra}{\la {\bf k}|a^{\dg}_R(x')a_R(x')|{\bf k}\ra},\nn 
\eea
where the denominators are a measure of the incident photon flux and are independent of $x,x'$. Here we choose $x>0$ and $x'<0$.
 Both $T$ and $R$ only provide information about average photon transport in the waveguide. Assessing other statistics of the scattered photons, such as fluctuations in the photon number, requires calculating higher-order correlation functions.

We shall now calculate the single-photon and two-photon scattering eigenstates for the Hamiltonian $H_{\rm eff}$ following \textcite{Roy10a, Zheng10}.

\begin{figure}[tb]
\includegraphics[height=4cm]{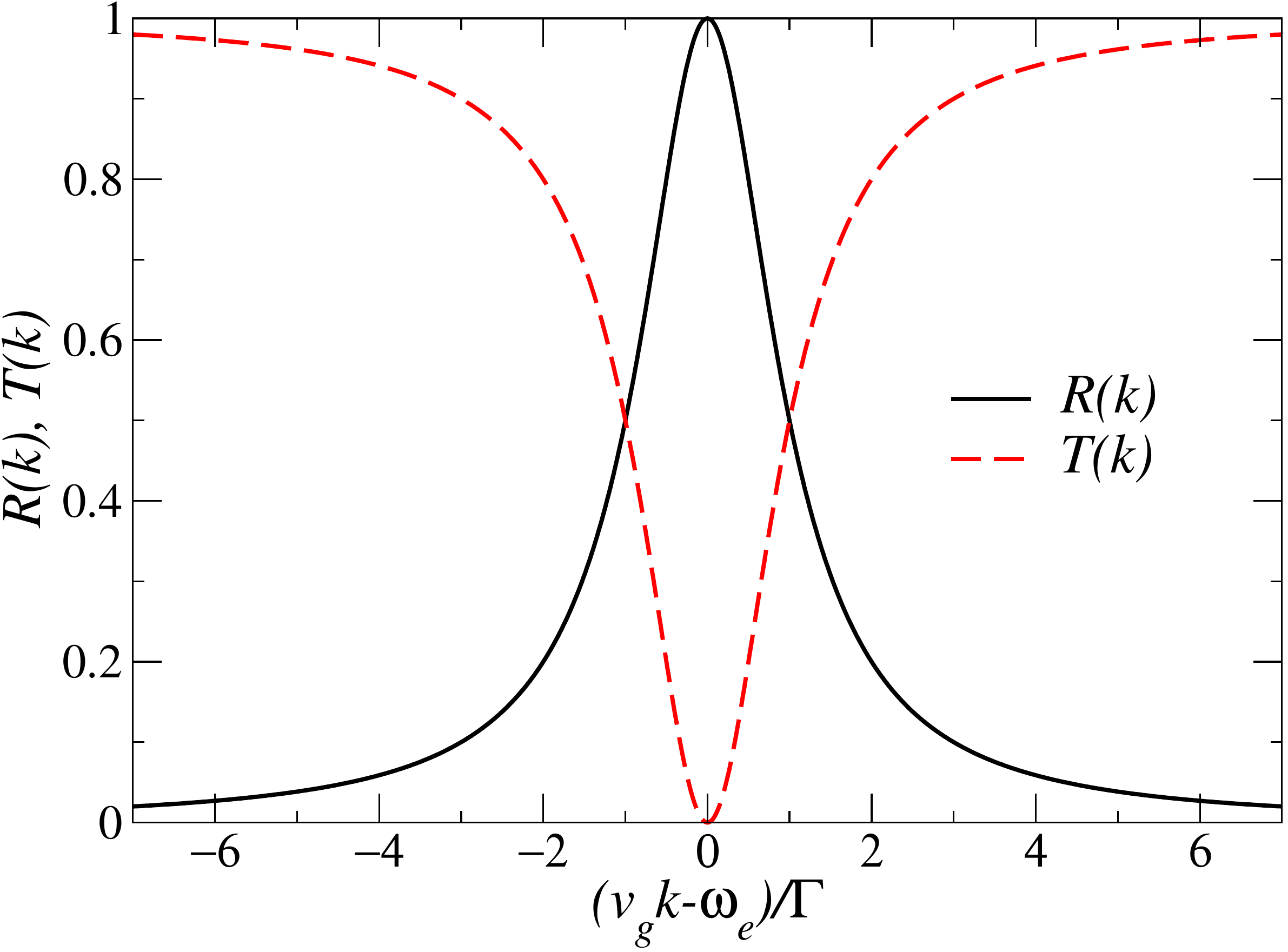}
\caption{(Color online). The reflection coefficient (solid line) and transmission coefficient (dashed line) of a single photon propagating in a 1D continuum with a side-coupled lossless two-level emitter.}
\label{spswitch}
\end{figure}

{\it Single-photon dynamics:} The states of an incident photon in the
right-moving channel is $|k\ra=\int dx e^{ikx} a^{\dg}_R(x)|\varphi \ra/\sqrt{2\pi}$,
where $|\varphi\ra$ represents the photon vacuum with the 2LE in the ground state. Considering different scattering
processes, we write an ansatz for the scattering eigenstate,
\bea
|k^+\ra&=&\int dx [g_R(x)a^{\dg}_R(x)+g_L(x)a^{\dg}_L(x)+\delta(x)e_k|e\ra \la g|]|\varphi\ra,\nn
\eea
where $g_R(x)$ and $g_L(x)$ are amplitudes for right-moving and
left-moving photons, respectively, and $e_k$ is the excitation amplitude
for the 2LE.  Using the Schr{\"o}dinger equation $H_{\rm eff}|k^+\ra=\hbar \vg k|k^+\ra$, we obtain three coupled linear equations for these three unknown amplitudes
\bea
\vg(-i\f{\partial}{\partial x}-k)g_R(x)+Ve_k\delta(x)&=&0,\nn \\
\vg(i\f{\partial}{\partial x}-k)g_L(x)+Ve_k\delta(x)&=&0, \label{eq_unknowns}\\
(\om_e-i\gamma-\vg k)e_k+V(g_R(0)+g_L(0))&=&0. \nn
\eea
Their solutions with the boundary conditions
$g_R(x<0)=e^{ikx}/\sqrt{2\pi}$, $g_L(x>0)=0$ and the continuity relation $g_{R/L}(0)=(g_{R/L}(0+)+g_{R/L}(0-))/2$ are
\bea
g_R(x)&\equiv&g_k(x)=\f{e^{ikx}}{\sqrt{2\pi}}\Big[\theta(-x)+t_k\theta(x)\Big], \nn\\
g_L(x)&=&\f{e^{-ikx}}{\sqrt{2\pi}}r_k\theta(-x), \label{SBA8}\\
e_k&=&\f{1}{\sqrt{2\pi}}\f{V}{\vg k-\omega_e+i(\gamma+\Gamma)}, \nn
\eea
where $\theta(x)$ is the step function. Here $\Gamma=V^2/\vg$, and we identify $2\Gamma$ as the energy relaxation rate of the emitter into the output channels.
The transmission amplitude $t_k$ and reflection amplitude $r_k$ are given by
\bea
t_k&=&\f{\vg k-\omega_e+i\gamma}{\vg k-\omega_e+i(\gamma+\Gamma)},~~r_k=t_k-1, \label{SBA11tk}
\eea
yielding the normalized 1-photon reflection and transmission coefficients
\bea
R(k)&=&|r_k|^2=\f{\Gamma^2}{(\vg k-\omega_e)^2+(\gamma+\Gamma)^2}, \label{SBA11}\\
T(k)&=&|t_k|^2=\f{(\vg k-\omega_e)^2+\gamma^2}{(\vg k-\omega_e)^2+(\gamma+\Gamma)^2}. \label{SPTr}
\eea
In the absence of loss ($\gamma=0$), $R(k)+T(k)=1$, and the 1-photon reflection exhibits a Breit-Wigner-like (Lorentzian) lineshape around the
resonance $\vg k=\omega_e$, as shown in Fig.~\ref{spswitch}. An incident, resonant photon is totally reflected by the emitter. Thus, a lossless side-coupled emitter behaves as a perfect mirror for propagating photons in a 1D continuum \cite{Shen05a, Shen05b}.

\begin{figure}[tb]
\includegraphics[height=5cm]{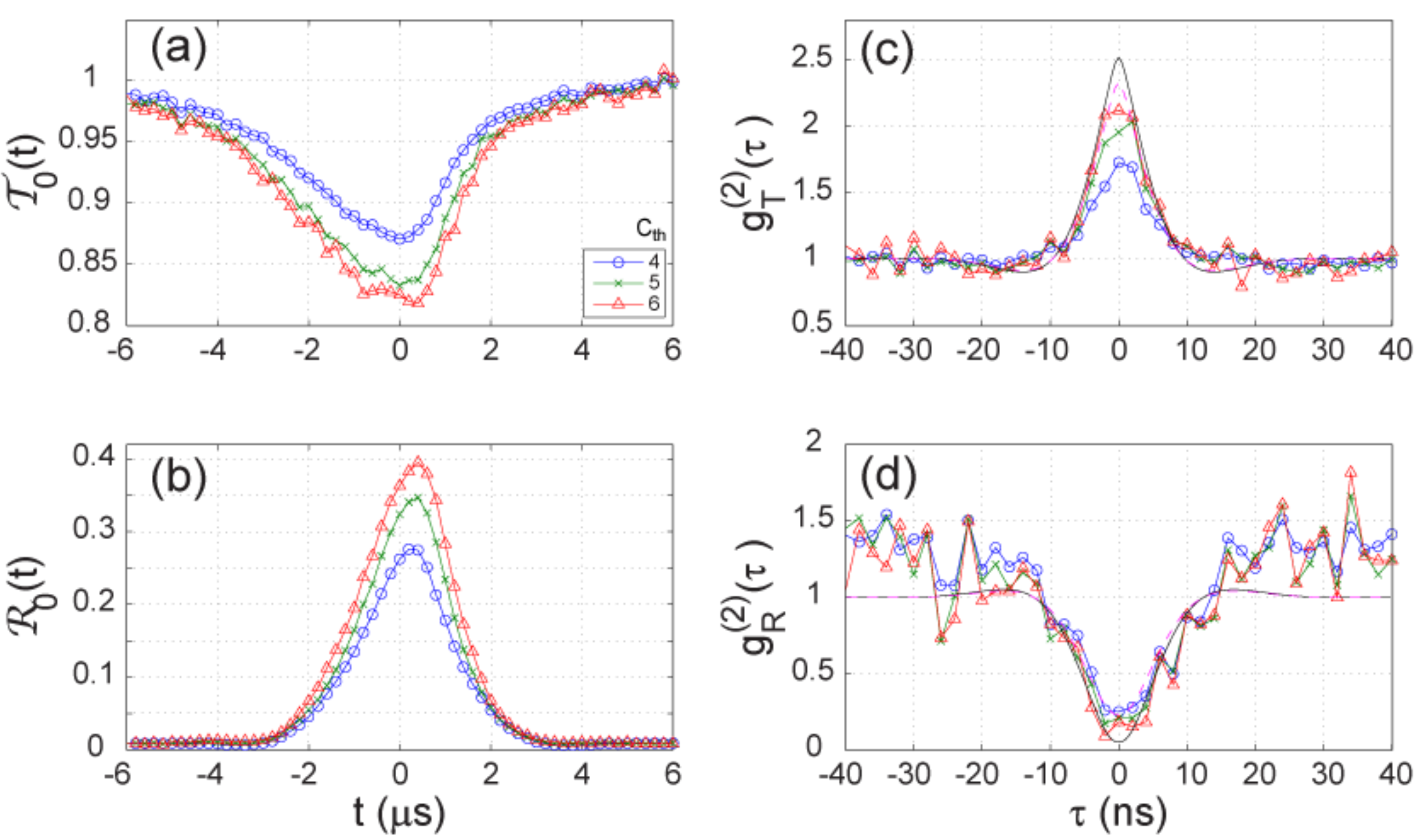}
\caption{(Color online). Reflection of single photons from a coherent-state input by one atom near a microtoroidal cavity. (a) Transmission  $\mathcal{T}_0(t)$ and (b) reflection
  $\mathcal{R}_0(t)$ of the probe field after averaging over atom transit
  events with a selection criterion that the sum of the counts over a time
  interval of 4 $\mu$s is equal to or greater than a threshold count $C_{\rm
    th}$ shown in the legend of (a). (c), (d) The second-order (intensity) correlation functions
  $g^{(2)}_{T,R}(\tau)$ for the transmitted and reflected fields. The dip in
  $g^{(2)}_{R}(\tau)$ around $\tau = 0$ indicates anticorrelation in the detection of photons, which is a signature of antibunching  and indicates that the field is dominated by single photons. Solid lines are a
  theoretical calculation described in \textcite{Aoki09} from which the figure
  is adapted.}
\label{aoki}
\end{figure}

Several features of these 1-photon lineshapes, including both the real and imaginary parts of $t_k$ and $r_k$, were observed in the optical regime in various \emph{cavity} QED setups \cite{Birnbaum05}. For example, a $40\%$ reflection of weak coherent light was obtained with microtoroidal cavities interacting
with single cesium atoms \cite{Dayan08,Aoki09}, as depicted in Figs.~\ref{aoki}(a) and (b).  The observation of photon antibunching (Fig.~\ref{aoki}(d)) in the reflected signal demonstrated that it was dominated by single photons. The coupling to and from the cavity was implemented
with a tapered optical fiber in the so-called overcoupled regime and thus dominated the internal system losses \cite{Aoki09}. The 1-photon lineshapes
were also demonstrated with a photonic crystal nanocavity coupled to
a semiconductor QD \cite{Englund07, Fushman08}. In a \emph{1D continuum}, strong
scattering of single photons by a single emitter was first observed for microwave photons in superconducting circuits by \textcite{Astafiev10a}. Similar scattering lineshapes were later observed for microwave and optical photons in various 1D settings \cite{Abdumalikov10, Hoi11, Goban14}. This
strong scattering of single photons has been exploited
to construct various all-optical quantum devices, such as a 1-photon quantum
switch \cite{Zhou08}, quantum memory, and quantum gates~\cite{Koshino10, Ciccarello12, Zheng13b, Rosenblum14}.

{\it Two-photon dynamics:} The dynamics of two photons strongly coupled to a single emitter in 1D is interesting and nontrivial. A 2LE is saturated by a single resonant photon, and a second photon in the waveguide cannot be absorbed by the excited emitter. \textcite{Rephaeli12} show that the outcome of the interaction depends on the spectral bandwidth of the second photon's wavepacket. If the extent of the wavepacket is much longer than the spontaneous emission lifetime
$(\Gamma+\gamma)^{-1}$ of the emitter, the excited emitter first decays to the ground state and then completely reflects the second photon. In the opposite limit of an extremely short wavepacket, the emitter-photon interaction is inhibited, and the second photon is fully transmitted.
In both limits, the first and second photons interact independently with the emitter and thus remain uncorrelated. However, for the intermediate regime, the second photon stimulates the relaxation of the emitter to the ground state. The two photons then leave the emitter simultaneously, becoming correlated. This stimulated emission in a 1D continuum is special,
as the emission enhancement cannot be entirely attributed to photon indistinguishability,
but largely results from the photon correlation generated by the emitter \cite{Rephaeli12}.

We shall construct a 2-photon scattering eigenstate in 1D that covers all of the
above limits. First, we write the incident state for two photons with wavevectors ${\bf k}=(k_1,k_2)$
in the right-moving channels
\bea
|{\bf k}\ra=\int dx_1 dx_2 \phi_{\bf k}(x_1,x_2)\f{1}{\sqrt{2}}a^{\dg}_R(x_1)a^{\dg}_R(x_2)|\varphi\ra, \label{SBA12}
\eea
where $\phi_{\bf k}(x_1,x_2)=(e^{ik_1x_1+ik_2x_2}+e^{ik_1x_2+ik_2x_1})/(2\sqrt{2}\pi)$. Our ansatz for the 2-photon scattering eigenstate is
\bea
|{\bf k}^+\ra&=&\int dx_1dx_2 \big[
g_{RR}(x_1,x_2)\f{1}{\sqrt{2}}a^{\dg}_R(x_1)a^{\dg}_R(x_2)\nn\\
&+&e_R(x_1)\delta(x_2)a^{\dg}_R(x_1)|e \ra \la g| \nn\\
&+&g_{RL}(x_1;x_2)a^{\dg}_R(x_1)a^{\dg}_L(x_2)   \nn\\
&+&e_L(x_2)\delta(x_1)a^{\dg}_L(x_2)| e\ra \la g| \nn\\
&+&g_{LL}(x_1,x_2)\f{1}{\sqrt{2}}a^{\dg}_L(x_1)a^{\dg}_L(x_2)\big]|\varphi\ra,
\eea
where $g_{RR}(x_1,x_2),~g_{RL}(x_1,x_2)$ and $g_{LL}(x_1,x_2)$ are 2-photon
amplitudes, and $e_R(x_1)$ and $e_L(x_2)$ are amplitudes of right- and left- moving photons with the 2LE in the excited state. These five unknown amplitudes can be found by solving five
coupled linear differential equations, obtained from the 2-photon stationary
Schr{\"o}dinger equation, $H_{\rm eff}|{\bf k}^+\ra=\hbar \vg(k_1+k_2)|{\bf k}^+\ra$. One can solve the differential equations with boundary conditions, $g_{RR}(x_1,x_2<0)=\phi_{\bf
  k}(x_1,x_2),~g_{LL}(x_1,x_2>0)=0,~g_{RL}(x_1<0;x_2>0)=0$ for the unscattered
state in Eq.~(\ref{SBA12}), and continuity
relations for the amplitudes, {\it e.g.}, $g_{RR}(0, x)=g_{RR}(x,0)= [g_{RR}(0+, x) + g_{RR}(0-, x)]/2$.
For example, the amplitudes of the right-moving photons in terms of
 $g_k(x)$ and $e_k$ in Eqs.~(\ref{SBA8}) are
\bea
&&g_{RR}(x_1,x_2)=\f{1}{\sqrt{2}}\big[g_{k_1}(x_1)g_{k_2}(x_2)+2\f{\Gamma}{\vg}e_{k_1}e_{k_2}e^{i(k_1+k_2)x_c}\nn \\&&e^{i(k_1+k_2-2\omega_e/\vg)x/2}e^{-(\gamma+\Gamma)x/\vg}\theta(x)\theta(x_2)\big]+(x_1 \leftrightarrow x_2),\nn\\\\
&&e_R(x_1)=\big(g_{k_1}(x_1)e_{k_2}+g_{k_2}(x_1)e_{k_1}\big)\nn \\&&~~~+2i\f{V}{\vg}e_{k_1}e_{k_2}e^{i(\vg k_1+\vg k_2-\omega_e+i\gamma+i\Gamma)x_1/\vg}\theta(x_1),
\eea
where $x_c=(x_1+x_2)/2$ and $x=x_1-x_2$. We observe that the second terms in
$g_{RR}(x_1,x_2)$ and $e_R(x_1)$ decay to zero with increasing
$|x_1-x_2|$ and $|x_1|$, respectively. These terms are regarded as \emph{2-photon bound states}.

Recently, the existence of two-particle bound states in the presence of a
localized interaction has been discussed in the contexts of both photon
\cite{Shen07a, Liao10} and electron transport \cite{Dhar08, Nishino11}. Because the interaction is spatially confined, energy and momentum can be exchanged and redistributed between the photons (with the constraint of fixed total energy), enabling the development of photon-photon correlations. The correlation strength depends on $\Gamma$. Experimental evidence for the 2-photon bound state appears in second-order correlation measurements and the transmission and reflection coefficients of the scattered fields \cite{Hoi12, Firstenberg13}, which we discuss later.

\begin{figure}[tb]
\includegraphics[width=8.5cm]{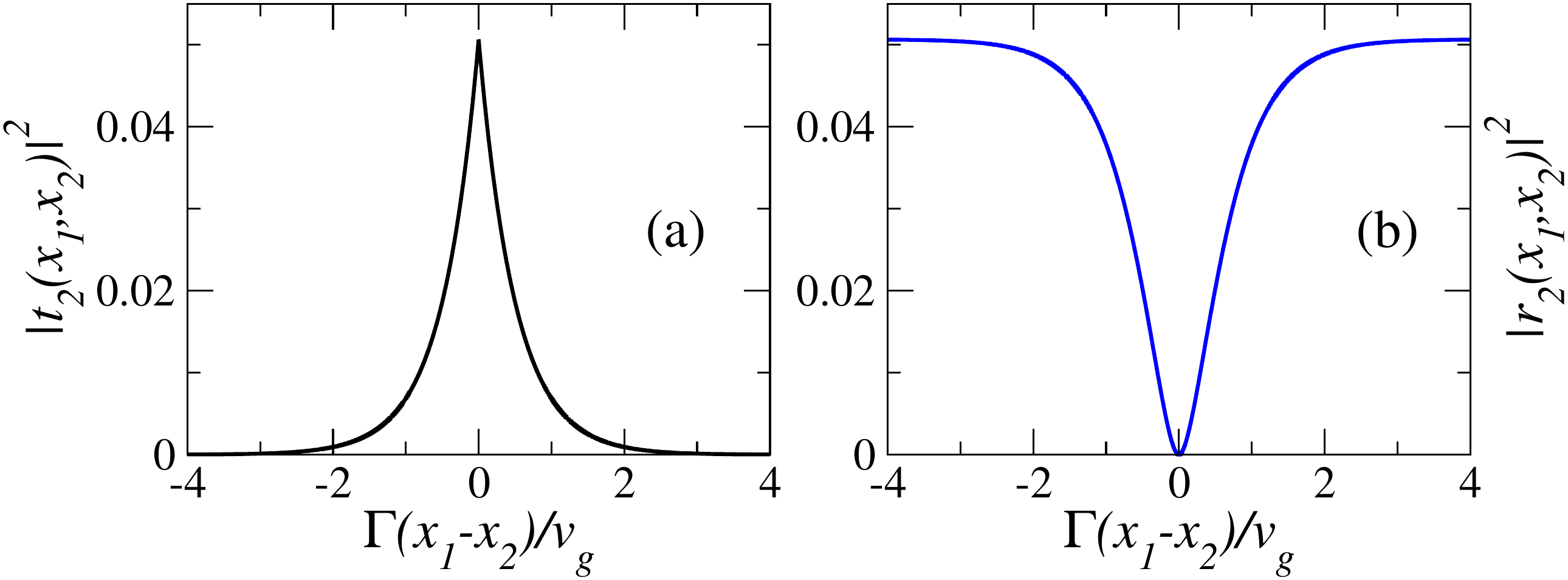}
\caption{(Color online). Correlations between scattered photons in the lossless resonant case.
The calculated two-photon scattering coefficients $|t_2|^2$ and $|r_2|^2$ are plotted versus the scaled separation $\Gamma(x_1-x_2)/\vg$.
(a) Bunching of transmitted photons indicated by the peak in $|t_2|^2$ at $x_1=x_2$. (b) Antibunching of reflected photons indicated by $|r_2|^2=0$ at $x_1=x_2$.
}
\label{2pcorr}
\end{figure}

It is interesting to study the asymptotic behavior (away from the emitter) of the 2-photon scattering eigenstate
\bea
&&|{\bf k}^{\rm a+} \ra=\int dx_1 dx_2
[rt(x_1,x_2)a^{\dg}_R(x_1)a^{\dg}_L(x_2)+ \nn\\&&
\f{r_2(x_1,x_2)}{\sqrt{2}}a^{\dg}_L(x_1)a^{\dg}_L(x_2)+\f{t_2(x_1,x_2)}{\sqrt{2}}a^{\dg}_R(x_1)a^{\dg}_R(x_2)]|\varphi \ra \nn,
\eea
where $t_2(x_1,x_2)$, $r_2(x_1,x_2)$ and $rt(x_1,x_2)$ are two-photon amplitudes where, respectively, both photons are transmitted, both are reflected, and one is transmitted and the other reflected. When two degenerate incident photons are resonant with the lossless emitter ($\vg k_1=\vg k_2=\omega_e$ and $\gamma=0$), we find \cite{Shen07b}
\bea
t_2(x_1,x_2)&=&-\f{1}{\sqrt{2}\pi}e^{2i\omega_e x_c/\vg}e^{-\Gamma|x|/\vg}, \\
r_2(x_1,x_2)&=&\f{1}{\sqrt{2}\pi}e^{-2i\omega_e x_c/\vg}(1-e^{-\Gamma|x|/\vg}),\\
rt(x_1,x_2)&=&-\f{1}{\pi}e^{i\omega_e x/\vg}e^{-2\Gamma|x_c|/\vg},
\eea
Two important points to notice from the above expressions
are: (1) the outgoing state is not a product state, and (2) the transmitted photons are {\it bunched}, whereas the reflected
photons are {\it antibunched}. Here, the separation $x$ between the two scattered photons is equivalent to a time delay between them. When plotted versus $x$ in Fig.~\ref{2pcorr}, $|t_2(x_1,x_2)|^2$ shows a cusp, and $|r_2(x_1,x_2)|^2$ shows a dip at $x=0$. For a
side-coupled 2LE, since $r_2(x_1,x_2)$ arises entirely from emission without any contribution from the incident photons, the
antibunching in $r_2(x_1,x_2)$ confirms that a single emitter cannot
simultaneously emit two photons. On the other hand, the behavior of
$t_2(x_1,x_2)$ and $rt(x_1,x_2)$ involves interference between the incident
and emitted photons. The behaviors of $r_2(x_1,x_2)$ and $t_2(x_1,x_2)$ in
Fig.~\ref{2pcorr} agree qualitatively with the experimentally measured 2-photon
correlations $g^{(2)}(\tau)$ of reflected and transmitted photons, shown in Figs.~\ref{aoki}(c,d) and \ref{hoi} for optical and microwave photons.

An exact multiphoton scattering state for 2LEs \cite{Zheng10} and multilevel emitters \cite{Roy14, Zheng12} has been derived, extending the above scattering theory. Multiphoton bound states appear in these multiphoton scattering states. The `brute force' technique we have used for constructing the scattering eigenstates becomes much more laborious with increasing photon number $N$, as the possible scattering configurations rapidly increase. An efficient method to describe the evolution of an arbitrary initial state of the present system has been developed by \textcite{Yudson08} using the algebraic Bethe ansatz. This elegant method avoids both the difficulty of following numerous configurations and the subtle problems of normalization and completeness of states. However, with increasing photon number it becomes very difficult to extract useful information from these exact multiphoton scattering states, whether calculated using the algebraic Bethe ansatz or  scattering theory. 

\begin{figure}[tb]
\includegraphics[width=8.5cm]{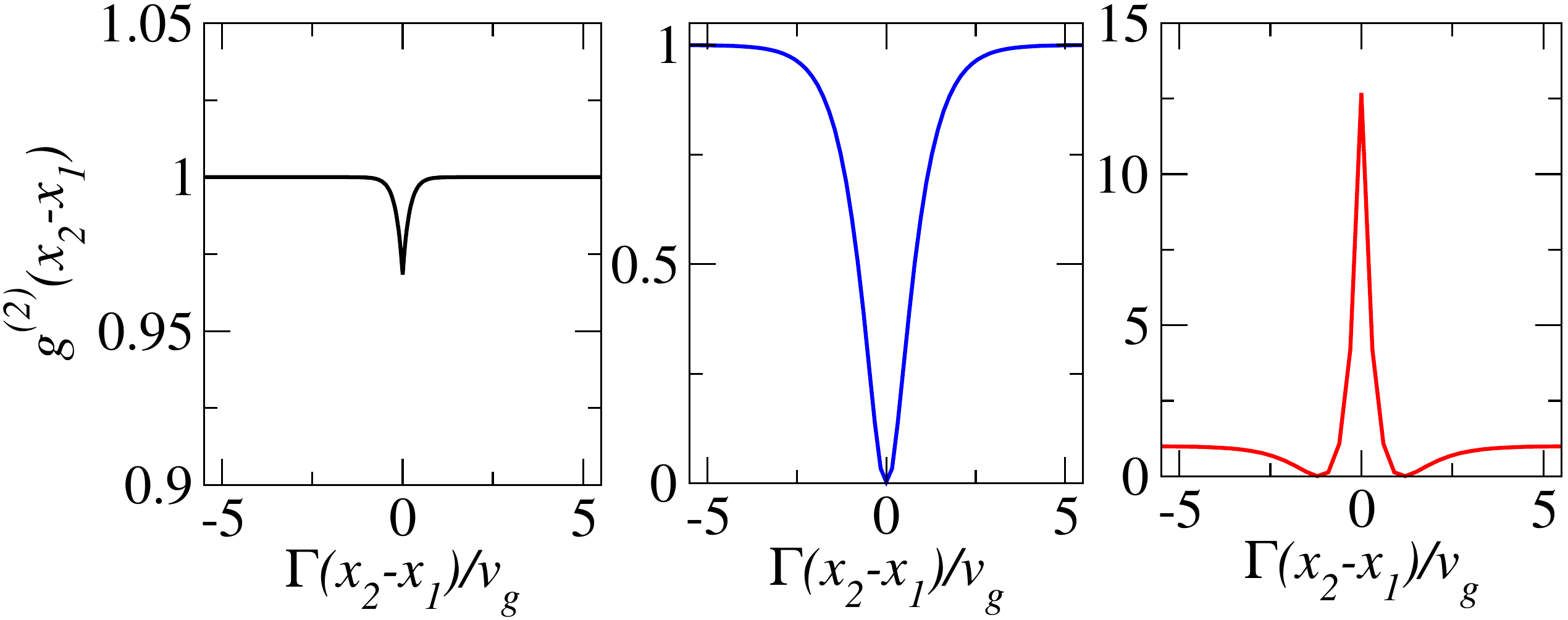}
\caption{(Color online). Second-order correlation function of the transmitted field at various emitter-photon coupling strengths for a coherent state input with $\bar{n}\le 1$. The couplings are (left)
$\Gamma/\gamma=0.2$, (middle) $\Gamma/\gamma=1.6$, and (right) $\Gamma/\gamma=3$. Here $\vg k_0=\om_e=10\gamma$ and $\Delta_k/\gamma=1$. Note the different vertical scales.
The second-order correlation shows antibunching and bunching at different coupling strengths.}
\label{2ndoc}
\end{figure}

{\it Coherent state input:} Coherent states are an important class of multiphoton states, describing the output of both an ideal laser and a microwave generator. They are thus commonly used as an input in experiments. \textcite{Zheng10} formulated the scattering of a weak coherent-state wave-packet by
a 2LE using the S-matrix in scattering theory. The incident wave-packet is
$|\alpha\ra=e^{a^{\dg}_{\alpha}-\bar{n}/2}|\varphi\ra$, where $\bar{n}=\int dk
|\alpha(k)|^2$ is the mean photon number, which is kept small $\bar{n} \le 1$
to observe few-photon behavior. For a coherent-state wave-packet, the photon statistics are Poissonian, implying that the variance of the photon number is also $\bar{n}$. Here $a^{\dg}_{\alpha}=\int dk \alpha(k)a^{\dg}_{R,k}$  for an incident wave-packet from the left. We employ a Gaussian wave-packet \cite{Zheng10}
\bea
\alpha(k)=\f{\sqrt{\bar{n}}}{(2\pi\Delta_k^2)^{1/4}}{\rm exp}\big(-\f{(k-k_0)^2}{4\Delta_k^2}\big),\nn
\eea
with width $\Delta_k$ and mean momentum $k_0$. The scattered state can be expressed as $|\psi^{\rm out}_{\alpha}\ra=\sum_{N}S^{(N)}|\alpha\ra,$
where $S^{(N)}$ is the S-matrix operator of $N$ incident photons, whose elements can be calculated using scattering theory. For example, $S^{(0)}=|\varphi\ra \la \varphi|$ and
\bea
S^{(1)}=\int dk~ t_k a^{\dg}_{R,k}|\varphi\ra \la \varphi|a_{R,k}+\int dk~ r_k
a^{\dg}_{L,k}|\varphi\ra \la \varphi|a_{R,k},\nn
\eea
where $t_k$ and $r_k$ are the 1-photon transmission and reflection amplitudes of Eq. (\ref{SBA11tk}).
Following this prescription, we find the transmission of a weak coherent state input
\bea
T(k_0,\Delta_k)&=&\f{\la \psi^{\rm out}_{\alpha} |a^{\dg}_{R}(x)a_R(x)| \psi^{\rm out}_{\alpha} \ra}{\la \alpha |a^{\dg}_{R}(x)a_{R}(x)| \alpha \ra}\nn\\
&=&\f{\int dk_1dk_2 \alpha (k_1) \alpha (k_2) t_{k_1}t^*_{k_2}
e^{i(k_1-k_2)x}}{\int dk_1 dk_2\alpha (k_1) \alpha (k_2)e^{i(k_1-k_2)x}},
\eea
where $x>0$, and we have kept only the leading order contribution coming from single-photon scattering. For a long, monochromatic pulse ($\Delta_k \ll \Gamma + \gamma$), the coefficient $T(k_0,\Delta_k)$ reduces to the single-photon transmission coefficient $T(k_0)$ in Eq.~(\ref{SPTr}).

In experiments, the statistics of scattered photons is predominately determined by measuring second-order correlations of the scattered fields. For the transmitted beam, it is defined as
\begin{equation*}
g^{(2)}(x_1,x_2)=\f{\la \psi^{\rm out}_{\alpha} |a^{\dg}_{R}(x_1)a^{\dg}_R(x_2)a_R(x_2)a_R(x_1)| \psi^{\rm out}_{\alpha} \ra}{\prod_{\substack{i=1}}^{2}\la \psi^{\rm out}_{\alpha}|a^{\dg}_{R}(x_i)a_R(x_i)|\psi^{\rm out}_{\alpha} \ra}.\label{2ndcoh}
\end{equation*}
By neglecting the contributions from $N\ge3$ photons in $|\psi^{\rm out}_{\alpha} \ra$ for $\bar{n} \le 1$, \textcite{Zheng10} have found
\bea
&&g^{(2)}(x_2-x_1)\label{2ndOC}\\
&&=\f{|\int dk_1dk_2\alpha(k_1)\alpha(k_2)(t_{k_1}t_{k_2}-r_{k_1}r_{k_2}e^{-(\Gamma+\gamma)|x|/\vg})|^2}{|\int dk_1dk_2 \alpha(k_1)\alpha(k_2)t_{k_1}t_{k_2}|^2}.\nn
\eea
Here again, the distance separation can be converted to time separation via
$\tau=(x_2-x_1)/\vg$, which is what is typically measured in experiments. The
first and second terms in the numerator of Eq.~(\ref{2ndOC}) are related
respectively to the noninteracting and the bound-state parts of the 2-photon
wavefunction. Without the bound-state contribution (\emph{i.e.}, no effective
photon-photon interaction), $g^{(2)}(x_2-x_1)=1$. In the presence of the bound
state, $g^{(2)}(x_2-x_1)$ can show {\it bunching} and {\it antibunching} of
the transmitted photons at different values of the coupling rate $\Gamma$, as
shown in Fig.~\ref{2ndoc}.   Note that the observability of these effects also
depends on having the appropriate value of $\Gamma/\gamma$. For very strong
coupling  $\Gamma/\gamma \gg 1$,  $g^{(2)}(x_2-x_1)$  of the
transmitted photons always shows bunching [Fig.~\ref{2ndoc}(c)] and was observed by \textcite{Hoi12}. For very weak coupling $\Gamma/\gamma \ll1$,  $g^{(2)}(x_2-x_1)$ is nearly featureless, exhibiting only a small antibunching dip [Fig.~\ref{2ndoc}(a)]. In the intermediate regime
$\Gamma/\gamma \approx 1$, the antibunching is more pronounced [Fig.~\ref{2ndoc}(b)]. 

\subsection{Input-output formalism}\label{sec-inout}
The input-output formalism is a celebrated technique for analyzing the effect of
light-matter interactions on the quantum statistics of light fields \cite{Gardiner85, Walls08}. Based on the Heisenberg picture, it allows one to study the time evolution of the field operators, with the ability to account for various input field states, for example, coherent states, Fock states or squeezed states. \textcite{Fan10} recently adopted this formalism to
investigate the few-photon scattering by emitters in a 1D continuum, relating it to the scattering theory of correlated photons.

{\it Equations of motion:} Essentially, given a Hamiltonian $H$, one uses the Heisenberg picture to derive a set of nonlinear differential equations for the time evolution of the input and output fields $b_{m,\rm in}(t),~b_{m,\rm out}(t)$ of left-moving ($m=L/+$) and right-moving ($m=R/-$) photons,
\bea
b_{m,\rm in}(t)&=&\f{1}{\sqrt{2\pi}}\int dk~ a_{m,k}(t_0)e^{i m \vg k(t-t_0)},\\
b_{m,\rm out}(t)&=&\f{1}{\sqrt{2\pi}}\int dk~ a_{m,k}(t_1)e^{i m \vg k(t-t_1)},
\eea
where $a_{m,k}(t_0)=e^{iHt_0/\hbar}a_{m,k}e^{-iHt_0/\hbar}$ and $a_{m,k}(t_1)=e^{iHt_1/\hbar}a_{m,k}e^{-iHt_1/\hbar}$  are Heisenberg operators in the limits $t_0 \to -\infty,~t_1 \to \infty$.  For clarity, we will use $b,b^{\dag}$ for operators in the input-output formalism and $a,a^{\dag}$ for operators in scattering theory. The input-output operators are directly related to the operators that create and destroy incoming and outgoing scattering eigenstates in scattering theory [see Eqs.~(\ref{stinput}) and (\ref{stoutput})]. For example,
\bea
&&b_{R,\rm in}(t)=\f{1}{\sqrt{2\pi}}\int dk
e^{i Ht_0/\hbar}a_{R,k}e^{-i Ht_0/\hbar}e^{-i\vg k(t-t_0)} \nn\\
&&=\f{1}{\sqrt{2\pi}}\int dk
e^{i Ht_0/\hbar}e^{-i H_0t_0/\hbar}a_{R,k}e^{iH_0t_0/\hbar}e^{-iHt_0/\hbar}e^{-i \vg kt} \nn\\
&&=\f{1}{\sqrt{2\pi}}\int dk a_{R,\rm in}(k)e^{-i \vg kt},\label{relation1}
\eea
where in the second line we replaced $a_{R,k}e^{i \vg kt_0}$ with $e^{-iH_0t_0/\hbar}a_{R,k}e^{iH_0t_0/\hbar}$ by employing $[H_0,a_{R,k}]=-\hbar \vg ka_{R,k}$. Thus, $a_{m,\rm in/out}(k)$
gives the spectral representation of $b_{m,\rm in/out}(t)$.  Using
relations similar to (\ref{relation1}), we can rewrite the S-matrix elements
in Eq.~(\ref{sminout}) as
\bea
&&S^{(N)}_{{\bf p};{\bf k}}=\la {\bf p}|S|{\bf k}\ra \\
&&=\la \varphi| a_{o_1,\rm out}(p_1)..a_{o_N,\rm out}(p_N)a^{\dg}_{i_1,\rm
  in}(k_1)..a^{\dg}_{i_N,\rm in}(k_N)|\varphi\ra \nn\\
&&=\mathcal{FT}^{2N}\la \varphi| b_{o_1,\rm out}(t_1)..b_{o_N,\rm out}(t_N)b^{\dg}_{i_1,\rm
  in}(t_1')..b^{\dg}_{i_N,\rm in}(t_N')|\varphi\ra, \nn
\eea
where  we use a global Fourier transform $\mathcal{FT}^{(2N)}=(2\pi)^{-N}\int \prod_{j=1}^N dt_jdt'_j e^{i\vg(i_jk_jt_j'-o_jp_jt_j)}$ with $i_j,o_j=-(R),+(L)$ to relate the S-matrix in linearized momentum (or frequency) and time. We shall now show how the S-matrix elements are found within the input-output formalism.

To simplify the presentation, we slightly rewrite the emitter part in the Hamiltonian
$H$ using Pauli matrices. We write the full Hamiltonian for a single 2LE
side-coupled to a 1D continuum as
\bea
&&\frac{H_{\rm io}}{\hbar}=\int_{-\infty}^{\infty} dk~
\vg k(a^{\dg}_{R,k}a_{R,k}-a^{\dg}_{L,k}a_{L,k})+\f{1}{2} \om_e
\sigma_z\nn\\
&&+\frac{V}{\sqrt{2\pi}}\int_{-\infty}^{\infty}
dk[(a^{\dg}_{R,k}+a^{\dg}_{L,k})\sigma_-+\sigma_+(a_{R,k}+a_{L,k})],\label{Hamio}
\eea
where $\sigma_{\pm}$ are raising and lowering operators for the 2LE and
$\sigma_z=2\sigma_+\sigma_--1$. Here $\omega_e$ is again the transition energy
of the 2LE, and we have dropped the inelastic loss term
$i\gamma$. We again assume that the coupling $V$ to the linearized modes is independent of the
wavevector $k$.

One can write Heisenberg equations of motion for the operators in Eq.~(\ref{Hamio}) and define input-output operators for
the fields, as illustrated in detail for a chiral model in
Appendix~\ref{inoutform}. One then gets
\bea
&&b_{R,\rm out}(t)=b_{R,\rm in}(t)-i\f{V}{\vg}\sigma_-(t),\nn\\
&&b_{L,\rm out}(t)=b_{L,\rm in}(t)-i\f{V}{\vg}\sigma_-(t),\label{io3}\\
&&\f{d\sigma_-}{dt}=-(i\omega_e+\Gamma)\sigma_-+iV\sigma_z[b_{R,\rm in}(t)+b_{L,\rm in}(t)],\nn
\eea
again denoting $\Gamma=V^2/\vg$. We now have all the required tools to study the scattering of photons in this system.

{\it Scattering properties:} The 1-photon scattering properties are encoded in the 1-photon S-matrix. For a single, right-moving input photon,
the transmission amplitude is given by the following element of the S-matrix
\bea
&&\la \varphi|  a_{R,\rm out}(p) a^{\dg}_{R,\rm in}(k)|\varphi\ra
\nn\\
&&=\f{1}{\sqrt{2\pi}}\int dt \la \varphi|b_{R,\rm out}(t)|k^+\ra e^{i \vg pt},\label{io4}
\eea
where we use $a^{\dg}_{R,\rm in}(k)|\varphi\ra=|k^+\ra$ [see
Eq.~(\ref{sminout})] and write $a_{R,\rm out}(p)$ in terms of $b_{R,\rm  out}(t)$. We therefore need to calculate $\la \varphi|b_{R,\rm out}(t)|k^+\ra$ to find the 1-photon transmission amplitude. It can be obtained by sandwiching
Eqs.~(\ref{io3}) between $\la \varphi|$ and $|k^+\ra$,
\bea
\la \varphi|b_{R/L,\rm out}(t)|k^+\ra&=&\la \varphi|b_{R/L,\rm
  in}(t)|k^+\ra\label{io5}\\
&-&i\f{V}{\vg}\la
\varphi|\sigma_-(t)|k^+\ra, \nn\\
\f{d}{dt}\la \varphi|\sigma_-|k^+\ra&=&-(i\omega_e+\Gamma)\la \varphi|\sigma_-|k^+\ra\label{io6}\\
&+&iV\la \varphi|\sigma_z[b_{R,\rm in}(t)+b_{L,\rm in}(t)]|k^+\ra.\nn
\eea
One can easily find some parts of the above expressions,
\bea
&&\la \varphi|b_{R,\rm in}(t)|k^+\ra=\la \varphi|b_{R,\rm in}(t)a^{\dg}_{R,\rm
  in}(k)|\varphi\ra=\f{e^{-i \vg kt}}{\sqrt{2\pi}},\nn\\
&&\la \varphi|b_{L,\rm in}(t)|k^+\ra=\la \varphi|b_{L,\rm in}(t)a^{\dg}_{R,\rm
  in}(k)|\varphi\ra=0,\label{io8}\\
&&\la \varphi|\sigma_zb_{R,\rm in}(t)|k^+\ra=-\la \varphi|b_{R,\rm
  in}(t)|k^+\ra,\nn
\eea
where $\sigma_z|\varphi\ra=-|\varphi\ra$, as $|\varphi\ra$ is the photon vacuum state with the 2LE also in its ground state.  Plugging the results of
Eqs.~(\ref{io8}) into Eq.~(\ref{io6}), we obtain a first-order
inhomogeneous differential equation with the solution
\bea
&&\la \varphi|\sigma_-|k^+\ra=\f{e^{-i \vg kt}}{\sqrt{2\pi}}\f{V}{\vg k-\omega_e+i\Gamma},\label{io10}\\
&&\la \varphi|b_{R,\rm out}(t)|k^+\ra=\f{\tilde{t}_ke^{-i \vg kt}}{\sqrt{2\pi}}.\label{io11}
\eea
Here $\tilde{t}_k = t_k(\gamma = 0)$ is the 1-photon transmission amplitude ignoring decoherence and photon loss. Finally, inserting Eq.~(\ref{io11}) in Eq.~(\ref{io4}), we arrive at
\bea
\la \varphi|  a_{R,\rm out}(p) a^{\dg}_{R,\rm in}(k)|\varphi\ra=\tilde{t}_k\delta(k-p).
\eea
Similarly, 
we can derive the 1-photon reflection amplitude $\tilde{r}_k=\tilde{t}_k-1$ 
from the S-matrix element $\la \varphi|  a_{L,\rm out}(p) a^{\dg}_{R,\rm in}(k)|\varphi\ra=\tilde{r}_k\delta(k+p)$. We thus see that the 1-photon scattering amplitudes obtained here are the same as those we found using scattering theory.

We note that the same result can also be arrived at by
directly approximating $\sigma_z=-1$ in Eqs.~(\ref{io3}), thus linearizing the operator equations. Such an approximation is commonly used in many quantum
optics calculations in the
weak-excitation limit by considering the 2LE to be mostly in its
ground state. Physically, the weak-excitation limit is valid for 1-photon
scattering when the 1-photon wavepacket has a much longer duration than the
 lifetime of the 2LE. However, the weak-excitation limit is not
always valid, even for a 1-photon pulse \cite{Rephaeli10}.

Similar derivations of the elements of the S-matrix are  carried out for multiphoton Fock states \cite{Fan10,Rephaeli11, Xu13} as well as for
coherent state inputs \cite{Koshino12, Peropadre13}. Input-output theory can also be
easily extended using the master equation formalism to include decoherence
processes such as pure dephasing, which dominates in superconducting systems
\cite{Koshino12, Peropadre13}. Finally, the input-output formalism has been recently extended to
investigate scattering of multiple photons by multiple interacting and
noninteracting emitters in a 1D continuum \cite{Xu15, Caneva15}. We discuss these in the next section.

\subsection{Other theoretical techniques}
\label{Otheory}
Until now we have discussed scattering of photons in a 1D continuum with a linearized dispersion relation and the Markov approximation. However, there are examples where the
linearization is a poor approximation, and the nonlinearity gives rise to important physical behavior \cite{Zhou08, Longo10, Roy11b}. One such case is of coupled resonator arrays, exhibiting a
tight-binding dispersion relation that is strongly nonlinear.
These structures were realized in photonic crystals \cite{Notomi08} and proposed in superconducting systems \cite{Zhou08}.

For a sinusoidal tight-binding dispersion, the Hamiltonian in Eqs.~(\ref{sc2LE}) and (\ref{sc2LEb}) can be rewritten as
\bea
\f{H_{\rm TB}}{\hbar}&=&-J\sum_{x=-\infty}^{\infty}(a_x^{\dg}a_{x+1}+a_{x+1}^{\dg}a_x) +(\omega_e-i\gamma)b^{\dg}b\nn\\
&+&V_0(a_{0}^{\dg}b+b^{\dg}a_{0})+\f{U}{2}~b^{\dg}b(b^{\dg}b-1),\label{HamTB}
\eea
where $a_x^{\dg}$ creates a photon at site
$x$, and $J$ is the hopping rate between nearest neighbor sites. Here, the 2LE is replaced by an additional bosonic site, side-coupled at $x=0$ 
\cite{Longo10,Roy11b}. The photon creation operator at the additional site is $b^{\dg}$. The states of 0 and 1 photon correspond respectively to the ground and the excited states of the 2LE in Eq.~(\ref{sc2LE}), whereas the forbidden `multiphoton' occupancy of the 2LE is avoided by introducing the interaction term $Ub^{\dg}b(b^{\dg}b-1)/2$ and taking 
the limit $U\to \infty$.

Scattering eigenstates for one and two photons can be derived exactly for the
Hamiltonian in Eq.~(\ref{HamTB}) using the Lippmann-Schwinger equation. \textcite{Roy11b} compares 1-photon and
2-photon transmission for linear and (nonlinear) tight-binding dispersion relations and points out the effect of
band edges on the 2-photon transmission.
Here again, the 2-photon scattering eigenstates exhibit bound states due to inelastic exchange of photons.
A unique feature of this approach is the ability to exactly calculate 2-photon scattering states for \emph{multiple} emitters separated by arbitrary distances \cite{Zheng13a}, similarly to the calculation of multiple quantum impurities \cite{Roy10b}. Furthermore, one can numerically study
2-photon scattering by \emph{multilevel} emitters. 

\textcite{Longo10} investigate the scattering of few-photon states in 1D lattice models using numerical time-dependent wave-packet evolution. This framework allows one to analyze both
the dynamics of multiphoton wave packets that interact
with the emitter and the dynamics of the emitter itself. Employing the approach of 
density matrix
renormalization group (DMRG), they extend their study for 3-photon and
4-photon transport. They demonstrate that a single-particle
photon-atom bound state with an energy outside the band can be excited via
multiparticle scattering processes, which leads to radiation trapping at the emitter. Another study worth noting is based
on the Lehmann-Symanzik-Zimmermann reduction, a method to calculate S-matrix
elements from the time-ordered correlation functions.  This approach
is employed by \textcite{Shi09} to investigate multiphoton
S-matrices in various complex quantum networks of propagating photons coupled
to emitters.

\subsection{Three-level emitter: a single-photon router}
\label{router}
A single-photon router can route one photon from an input port to
either of two output ports,
while conserving the
superposition of input photonic states. It thus finds important applications in optical quantum networks. 
Building on the work of \textcite{Abdumalikov10}, \textcite{Hoi11} demonstrated that a driven
three-level emitter (3LE) strongly coupled to a 1D continuum can act as an
efficient single-photon router \cite{Chang07, Neumeier13, Shomroni14}.
As we shall see, the presence or absence of a classical control field in this system determines a specific output port for the probe photon.

\begin{figure}[tb]
\includegraphics[trim={3cm 3cm 10cm 0.5cm},clip,width=6.0cm]{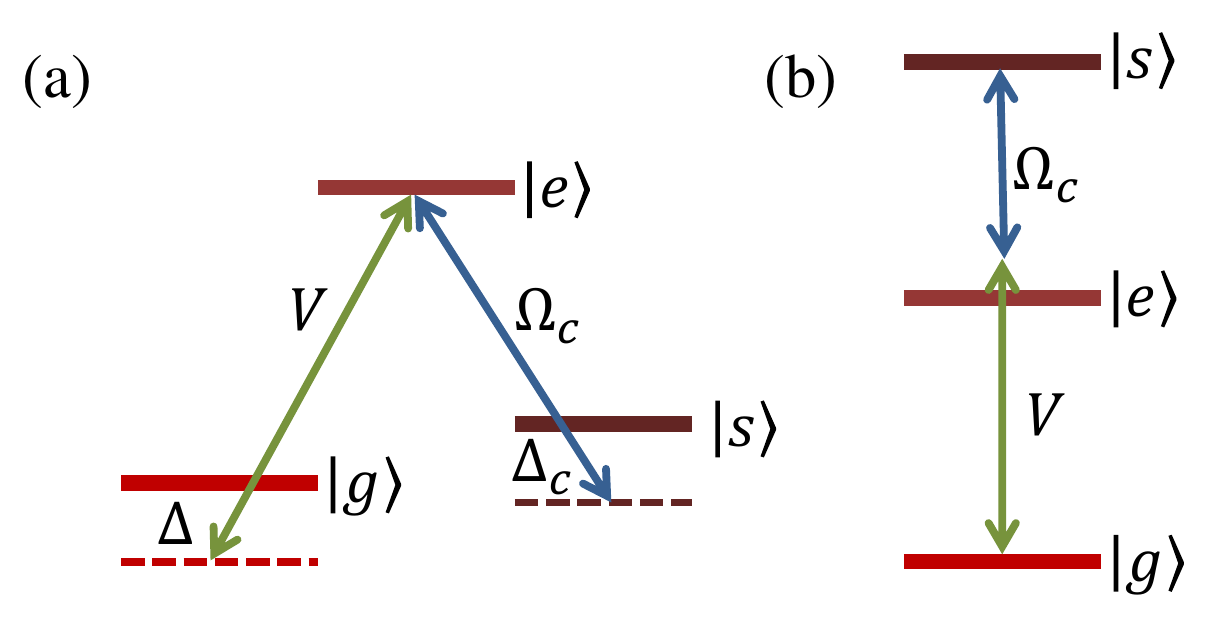}
\caption{(Color online). (a) $\Lambda$-type and (b) ladder-type three-level emitters.}
\label{sch}
\end{figure}

We consider a 3LE with either the $\Lambda$ or ladder-type structures in Fig.~\ref{sch}, with the probe light tuned to the $|g\ra-|e\ra$ transition. A classical light field drives the $|e\ra-|s\ra$ transition with a Rabi frequency $\Omega_c$ and frequency detuning $\Delta_c$. The Hamiltonian describing the system 
\bea
\frac{H_{\rm R}}{\hbar}=\frac{H_{\rm eff}}{\hbar}+(\om_e+\Delta_c-i\gamma_s)| s \ra\la s|+\frac{\Omega_c}{2}(| s \ra\la e|+| e \ra\la s|),\nn\\\label{HamSR}
\eea
extends the Hamiltonian $H_{\rm eff}$ of the 2LE in Eq.~(\ref{sc2LE1}).
Losses from the state $|s \ra$ are accounted for by the imaginary term $-i\gamma_s$.
With natural atoms, $|g \ra$ and $|s \ra$ can be two different Zeeman states, and the transitions $|g\ra-|e\ra$ and $|e\ra - |s\ra$ would couple to different optical polarizations based on selection rules.

\begin{figure}[tb]
\includegraphics[width=8.5cm]{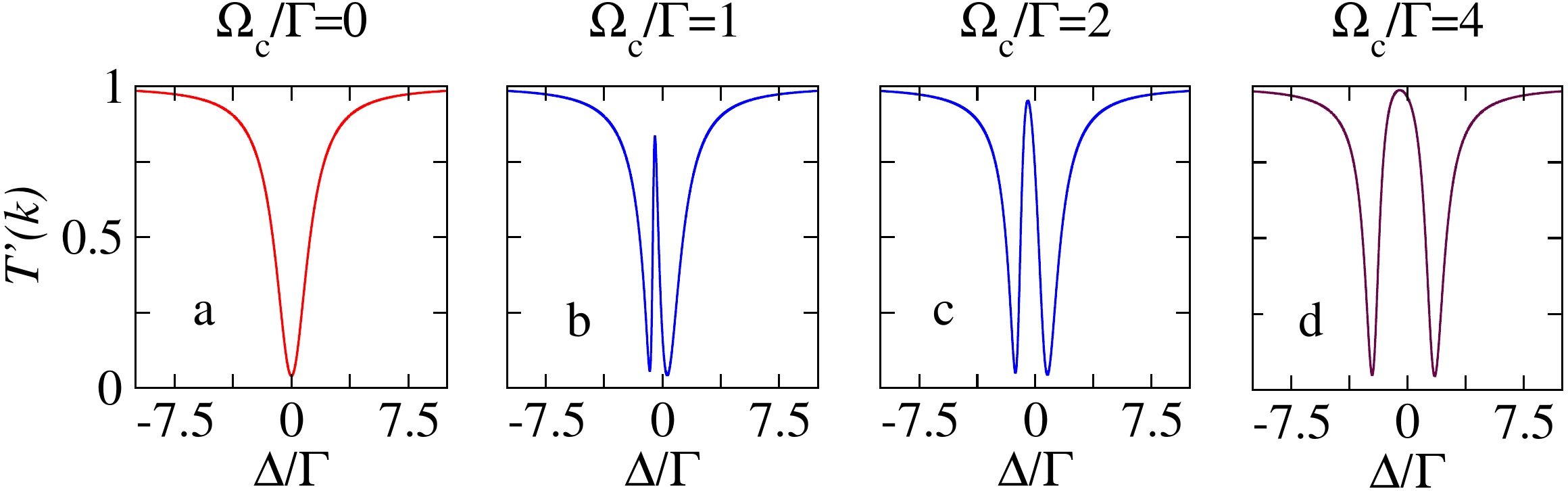}
\caption{(Color online). Prefect reflection (a), electromagnetically induced transparency at  the Raman resonance $\Delta=\Delta_c$ for weak control field (b,c)
and Autler-Townes splitting for stronger control fields (d). 
The splitting between the Autler-Townes doublet is $\Omega_c$. The parameters are
 $\Delta_c/\Gamma=-1/2,~\gamma/\Gamma=1/4,~\g_s/\Gamma=1/40$.}
\label{EIT1}
\end{figure}

\textcite{Witthaut10} and \textcite{Roy11a} study the transmission and reflection line-shapes for the $\Lambda$-type system. The
1-photon transmission and reflection amplitudes are given respectively
by $t_{k}'=
\chi/(\chi+i\Gamma)$ and
$r_{k}'=
-i\Gamma/(\chi+i\Gamma)$ [see, {\it e.g.}, \textcite{Roy14} for a derivation], where again $\Gamma=V^2/\vg$ and
\bea
&&\chi=\Delta+i\gamma-\f{\Omega_c^2}{4(\delta+i\gamma_s)}.\label{trans}
 \eea
Here $\Delta=\vg k-\om_e$ is the detuning of the incident probe
photon from the $|g \ra-| e\ra$ transition, and $\delta=\Delta-\Delta_c$ is the Raman detuning. 
We plot in Fig.~\ref{EIT1} the transmission coefficient
$T'(k)=|t_{k}'|^2$ for different values of the parameters. 
In the absence of the control field, the probe photon is reflected due to the $|g\ra-|e\ra$ transition [Fig.~\ref{EIT1}(a)], whereas
in the presence of a control, when $\Omega_c^2\gtrsim \Gamma \gamma_s$, a transmission window appears at the Raman resonance $\Delta=\Delta_c$ ($\delta=0$) [Figs.~\ref{EIT1}(b-d)].

Two parameter regimes are of interest. In the first regime, predominantly characterizing atomic $\Lambda$-systems, the $|s\ra$ state is metastable and much longer lived than the $| e\ra$ state, that is $\gamma_s\ll\gamma,\Gamma$. This yields a narrow transmission window within the broader reflection or absorption line, an effect known as electromagnetically induced transparency (EIT) \cite{Harris90, Boller91, Fleischhauer05}. Intuitively, EIT results from a destructive interference between two allowed transitions, leading to cancellation of the population of $|e\ra$ and to formation of a
`dark state'.  With increasing strength of the control field, the width of the transparency window increases, as shown in Fig.~\ref{EIT1}(c).

In the second regime, generally characterizing superconducting ladder systems, $|s\ra$ is shorter lived than $| e\ra$ due to population relaxation, as the $|s\ra-|e\ra$ transition also couples strongly to the transmission line. Therefore $\gamma_s \gtrsim \gamma,\Gamma$, and narrow EIT lines cannot be obtained. Rather, stronger control fields $\Omega_c \gg \Gamma$ are needed to drive the $|s\ra-|e\ra$ transition and open a transparency window \cite{Anisimov11}, an effect known as the Autler-Townes splitting (ATS). The width of the transmission window due to ATS is $\Omega_c$, as shown in Fig.~\ref{EIT1}(d). In both regimes, a sufficiently strong control field allows the probe photons to pass the emitter without being reflected. Thus, the presence or absence of a control field determines the route of the probe photon.

\begin{figure}[tb]
\includegraphics[width=8.5cm]{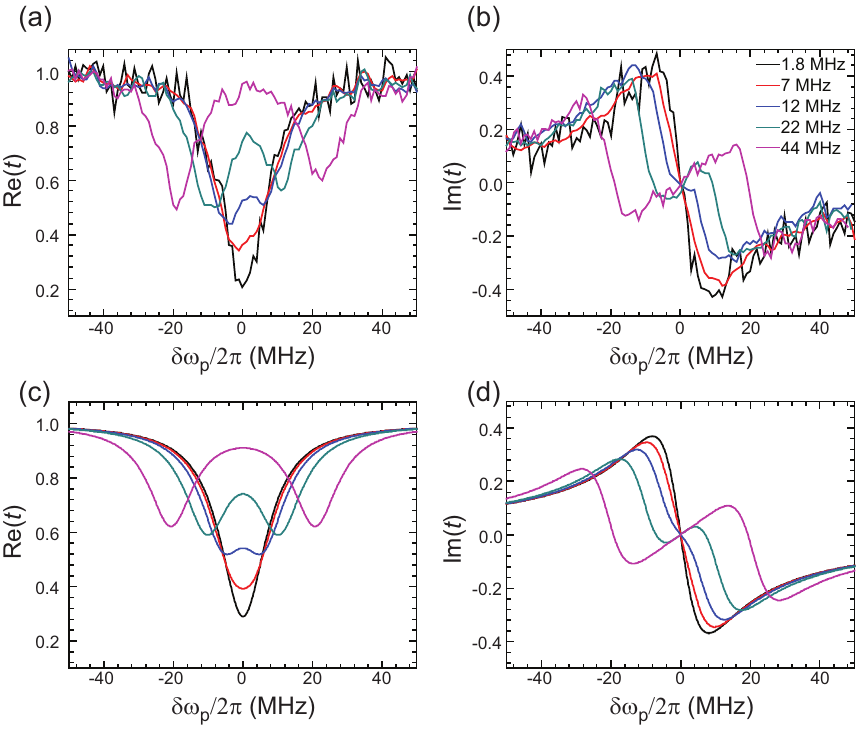}
\caption{(Color online). One-photon transmission amplitudes $t$ versus the probe detuning $\delta\omega_p\equiv\Delta$ for a superconducting three-level qubit coupled to a microwave waveguide. (a) and (b) are the experimentally measured
real and imaginary parts of the transmission amplitudes for various control field amplitudes $\Omega_c$, specified in (b). The curves show
typical dispersion for the Autler-Townes splitting. Re$(t)$ near $\delta\omega_p=0$ approaches unity with increasing $\Omega_c$. (c) and (d) are corresponding theoretical curves. Adapted from \textcite{Abdumalikov10}.}
\label{Abduma}
\end{figure}

The recent experiments by \textcite{Abdumalikov10,Hoi11} used a ladder-type superconducting qubit and observed transparency windows due to ATS, as shown in Figs.~\ref{Abduma}(a,b). By tuning the control field, \textcite{Hoi11} demonstrated the routing of a single-photon probe. Since both relaxation and dephasing are important in superconducting systems, a generalized model was introduced by \textcite{Abdumalikov10} to describe the experiments, replacing the effective non-Hermitian Hamiltonian (\ref{HamSR}) with a Markovian master equation for the density matrix. The transmission spectra calculated by \textcite{Abdumalikov10} are shown in Figs.~\ref{Abduma}(c,d) and are similar to our $t_{k}'$ given above [with their loss terms $\g_{21}$ and $\g_{31}$ corresponding roughly to our $(\gamma+\Gamma)$ and $\g_s$, respectively].

Recently, more complicated superconducting systems have been used to make effective $\Lambda$-systems. \textcite{Inomata14} realized an effective ``impedance-matched'' $\Lambda$-type emitter using dressed states of a driven superconducting qubit-resonator system. \textcite{Novikov2016}  demonstrated EIT in a superconducting circuit using the Jaynes-Cummings dressed states of a strongly coupled qubit-cavity system as an effective $\Lambda$-system.

Photon-photon correlations have also been calculated for 3LE systems \cite{Li15}. To do this, one needs to derive the multiphoton scattering states for the system. \textcite{Roy11a} derive the 2-photon scattering state of the probe field for a 3LE weakly driven by a control field. An exact 2-photon scattering state for an arbitrary strength of $\Omega_c$ is calculated by \textcite{Zheng12, Roy14}. \textcite{Roy14} show that the second-order correlation of the transmitted photons near the Raman resonance changes from bunching to antibunching to constant, as the strength of the control field is ramped up from zero to a higher value where the ATS appears.

\subsection{Superconducting circuits}
\label{SCcircuit}
While proposed in a variety of experimental systems, the strong coupling of propagating light to matter was first demonstrated in a circuit QED setting, with superconducting qubits playing the role of artificial atoms \cite{Astafiev10a}. The major advantage of these types of systems in achieving strong coupling is that the electromagnetic fields can be tightly confined into quasi-1D superconducting waveguides ($v_g \sim c/3$).
In a typical implementation, the lateral dimensions are of order 10 microns while the corresponding wavelength is of order 10 mm. (In the direction of propagation, the light is unconfined.) The mode volume is then $\sim 10^{-6} \lambda^3$, compared to $\sim 1 \lambda^3$ for optical systems, implying strongly enhanced electric field strengths.  It is, furthermore, straightforward to fabricate superconducting qubits with dipole moments in this range \cite{Wallraff04}, meaning that efficient mode matching can be easily accomplished. These were the key insights leading to the rapid growth of experimental work in this field.

These systems work in the microwave regime. 
In this frequency range ($\sim 5$ GHz), where the photon energy $\hbar \omega$ is well below the superconducting gap energy $\Delta_\textrm{SC}$, superconductors have very little loss, which protects the coherence of the circuits. In aluminum for instance, $\Delta_\textrm{SC}/h \approx 50$ GHz. At optical frequencies ($\hbar \omega \gg \Delta_\textrm{SC}$), metals are not superconductors and are very lossy.  Therefore, metallic waveguides are not useful in confining optical light at these size scales. The trade-off is that these microwave systems must be operated at very low temperatures, typically below 50 mK, in order for the background thermal (blackbody) field to be sufficiently suppressed.

\begin{figure}[tb]
\includegraphics[width=6cm]{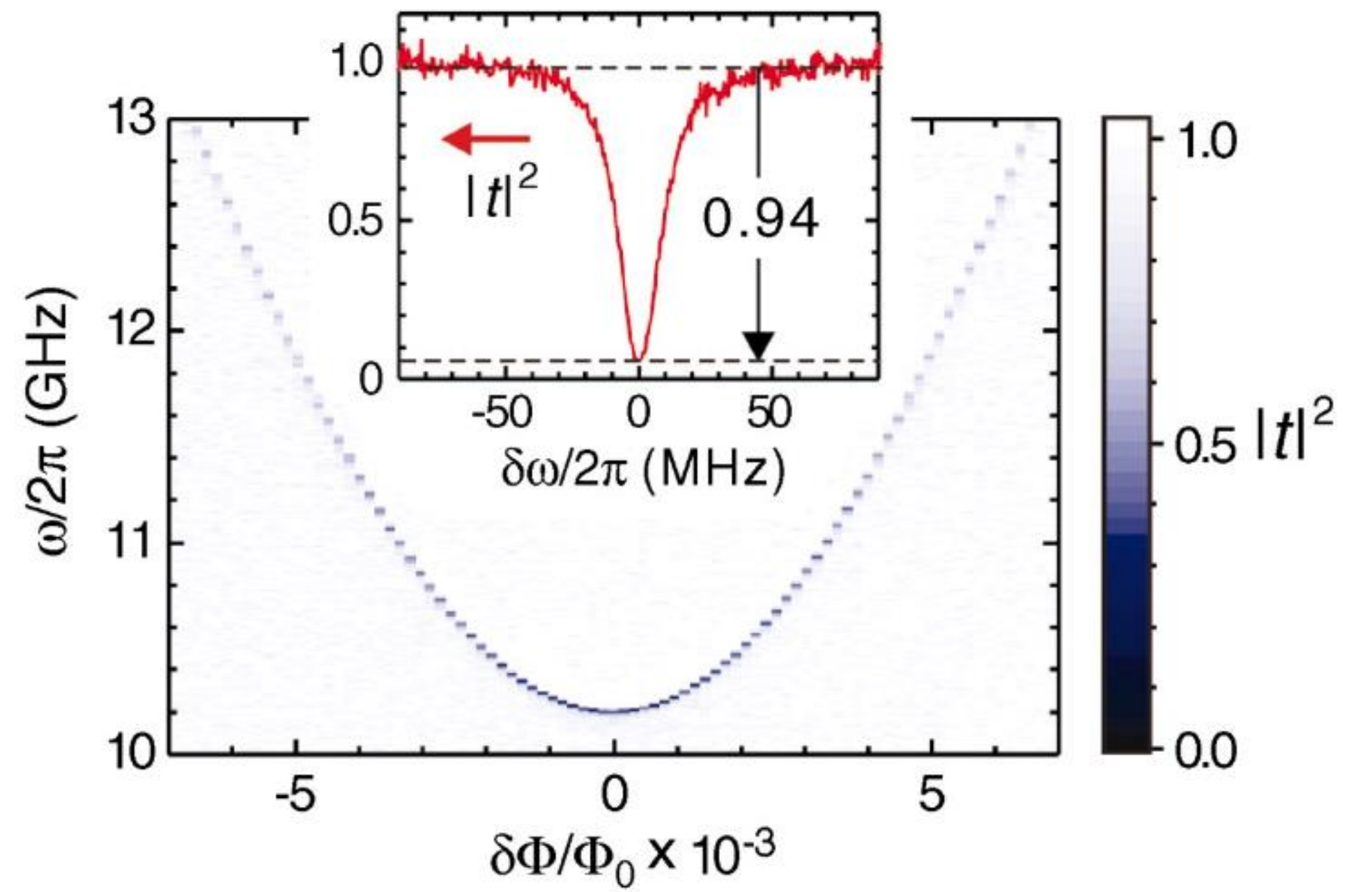}
\caption{(Color online). Spectroscopy of an artificial atom coupled to an open 1D transmission line. The colormap shows the power transmission $T=|t|^2$ versus flux bias $\delta \Phi$ and incident microwave frequency $\om$. When the incident radiation is on resonance with the emitter, a dip in $T$  reveals a dark line.
Inset: $|t|^2$ at $\delta \Phi=0$ as a function of the probe detuning $\delta\om\equiv\Delta$ from the resonance frequency $\om_e/2\pi=$10.204 GHz. The maximal power extinction of $1-T=94\%$ takes place on resonance ($\delta\om=0$). Adapted from \textcite{Astafiev10a}.}
\label{astafiev}
\end{figure}

The first demonstration of the strong scattering of propagating microwave
light by a single artificial atom was achieved by \textcite{Astafiev10a} from
the NEC Corporation group (see Fig.~\ref{astafiev}).  They observed a strong extinction of the transmitted light corresponding to $1-T = 94\%$. This was the first time that the hallmark result of
$1-T> 50\%$ was achieved for a single scatterer in any type of system, clearly
separating the coherent and incoherent scattering regimes. They also derived reflection $(r)$ and transmission $(t)$ amplitudes of an incident coherent state using input-output theory and the master equation approach to treat decoherence:
\bea
\label{SQCr}
r=r_0\f{1+i\Delta/\Gamma_2}{1+(\Delta/\Gamma_2)^2+\Omega^2/\Gamma_1\Gamma_2},~~t=1-r,
\eea
with $r_0$ the maximal reflection amplitude and $\Omega$ the Rabi frequency of the incident probe. The rate $\Gamma_2=\Gamma_1/2+\Gamma_{\varphi}$, where $\Gamma_1$ is the energy relaxation rate and $\Gamma_{\varphi}$ is the pure dephasing rate. Note that there is a sign difference in the definition of reflection amplitude in Eq.~\ref{SBA8} and \textcite{Astafiev10a}. For superconducting systems, the rate of emission into spurious modes is negligible, implying $\Gamma_1 = 2\Gamma$.  For weak driving ($\Omega \ll \Gamma_1,\Gamma_2$), Eq.~(\ref{SQCr}) agrees with Eq.~(\ref{SBA11tk}) if we further identify $\Gamma_{\varphi} = \gamma$.  We note again that this is a typical distinction between atomic and superconducting systems, that is, in the superconducting systems we consider here, emission into spurious modes is negligible while pure dephasing is significant.

The NEC group demonstrated a number of prototypical
atomic physics effects using their artificial atom including resonance
fluorescence of the Mollow triplet \cite{Astafiev10a, Astafiev10b} and induced transparency due to the Autler-Townes splitting
\cite{Abdumalikov10} as discussed before.  These results demonstrated that the quality of coherence in superconducting qubits had become sufficiently high for them to genuinely be considered ``artificial atoms."  NEC's work was reproduced by the Chalmers group, who improved the scattering efficiency by more than an order of magnitude by increasing the coupling strength, achieving $1-T = 99.4\%$ \cite{Hoi11, Hoi13a}.

Importantly, the simple measurement of the coherent scattering properties at single-photon probe powers can be
fully explained by a classical model of scattering from a harmonic oscillator,
\textit{i.e.}, an LC circuit.  Essentially, as long as the probe only
weakly excites the atom from the ground state, it does not obtain information
about the presence of higher levels and cannot distinguish between a two-level
and a multilevel system.
To rule out that the observed signal is purely classical, we must look beyond the linear response of the system.
The experiments mentioned above which involve stronger probes and nonlinear
response, such as the resonance fluorescence experiment, are not simply explained by a
classical model of the emitter.  Still, they tell us very little 
about the state of the scattered electromagnetic field.

\begin{figure}[tb]
\includegraphics[width=9.0cm]{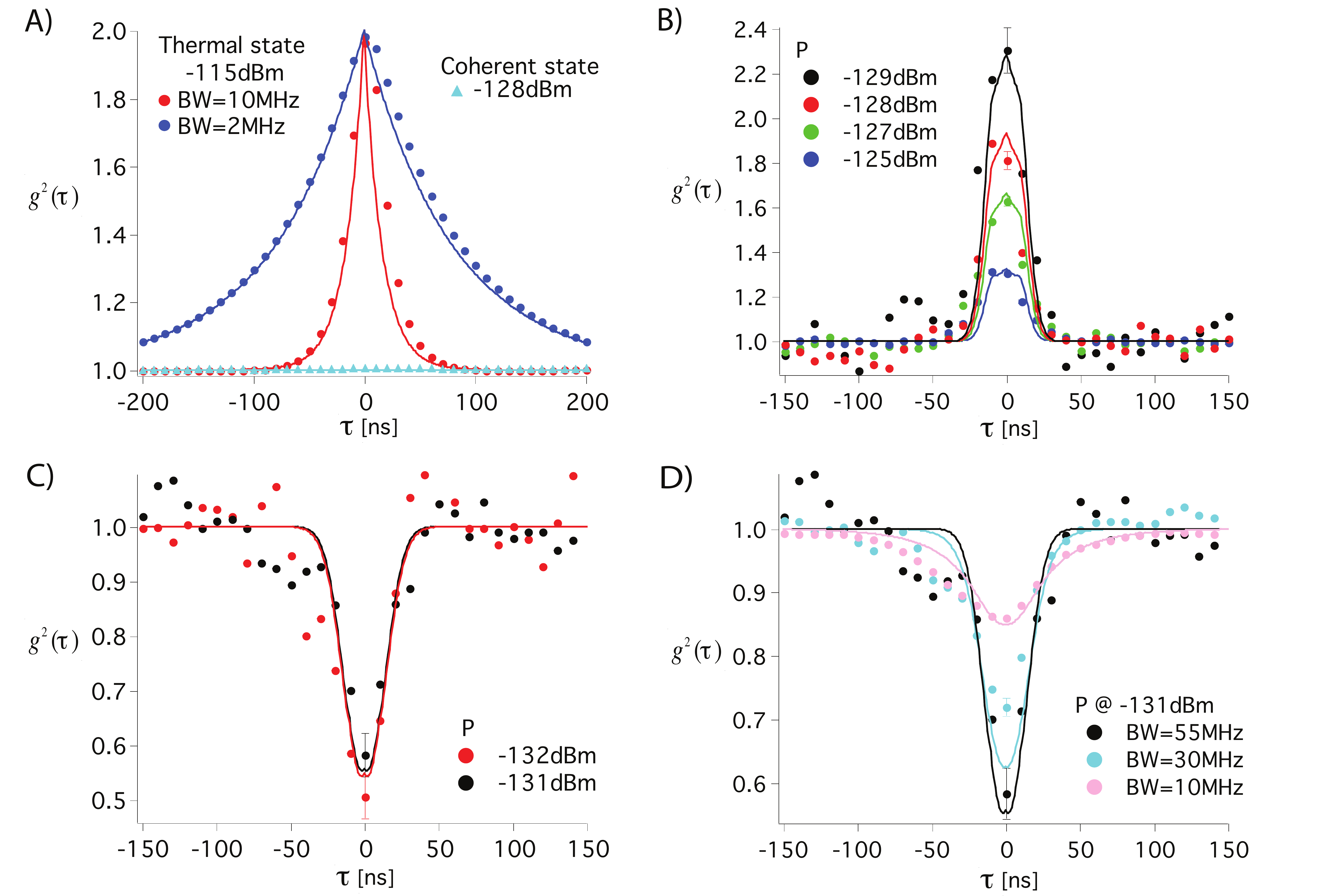}
\caption{(Color online). Second-order correlation function $g^{(2)}$ versus the time separation $\tau$ between the photons for a thermal state, a coherent
  state, and the scattered states generated by the artificial two-level
  emitter. (A) $g^{(2)}(\tau)$ of a thermal state and a coherent state. (B) $g^{(2)}(\tau)$ of the resonant
  transmitted microwaves for four different incident powers. (C) $g^{(2)}(\tau)$ of a resonant reflected field for two different incident powers. $g^{(2)}(0)$ does not reach zero because of well-understood experimental imperfections, including the bandwidth of the measurement system.  (D) $g^{(2)}(\tau)$ of a resonant reflected field for different measurement bandwidths (BW), illustrating how decreasing bandwidth decreases the depth of the dip.
 The solid curves in (A)-(D) are theory curves.
 Adapted from \textcite{Hoi12}.}
\label{hoi}
\end{figure}

As we have shown theoretically, the light scattered by a single emitter should be distinctly nonclassical, which can be characterized by higher-order correlation measurements.
The Chalmers group 
measured the second-order correlation function
$g^{(2)}(\tau)$ of the scattered field and showed that it was significantly
less than 1 \cite{Hoi12}.  The essential idea is that the emitter can only scatter one photon
at a time, so it only reflects the 1-photon component of the input coherent
state, while transmitting the higher photon number components.  The reflected
state is then a superposition of only the vacuum and the
1-photon state, which exhibits photon antibunching as  shown in
Fig.~\ref{hoi}. The antibunching behavior reveals the quantum nature of the
scattered field \cite{Kimble77, Paul82}.

These circuits have also been explored in terms of possible applications for
quantum communication networks.  One broad architecture of a 
network imagines quantum nodes, which perform basic
processing tasks, connected by long-distance channels carrying quantum
information \cite{Kimble08}. By far, the leading
candidates for implementing quantum channels involve optical or telecom
photons propagating in fibers or free space.  A number of physical systems are
being actively investigated for implementing quantum nodes \cite{Greve12,
  Gao12, Dicarlo09, Ritter12, Sherson06} with superconducting circuits in open environment being one of the promising candidates.

Using the architecture described above, the Chalmers group has already demonstrated a number of prototype elements for quantum nodes. In the first experiment, as mentioned earlier, they demonstrated a router that exploited the ATS to direct a microwave input at the single-photon level between two ports \cite{Hoi11}.  The on-off ratio of the router was 99\% with a switching time of a few nanoseconds.  The ability to produce nonclassical light, which is a required resource for a quantum network, by scattering classical light from a purely passive device as described above \cite{Hoi12} also has potential technological applications.
This can be considered as a ``quantum-state filter" that accepts a desired portion of the input state (the 1-photon component) and rejects the rest (the higher photon numbers).  We compare this to a conventional frequency filter that accepts a desired set of frequencies while rejecting others.  More advanced quantum-state filters, containing multiple qubits, could be an economical way to produce nonclassical light for quantum networks.

In a separate experiment, the Chalmers group demonstrated that the single
qubit worked as a highly effective cross-Kerr medium \cite{Hoi13}. The cross-Kerr effect is
essentially an effective interaction between light at two different
frequencies mediated by a nonlinear medium \cite{Shen1984}. The basic effect is that the
presence of one beam induces a phase shift in the other beam that is
proportional to intensity. The cross-Kerr effect has been studied extensively
in the past as a possible route to, \textit{e.g}, quantum nondemolition
measurements (QND) of single photons \cite{Munro2005} and photonic gates \cite{Turchette1995,Milburn1989}.  However, the strength of
the cross-Kerr effect is generally too weak in bulk nonlinear crystal for any
of these applications to be realizable. In the Chalmers experiment, by contrast, the bulk nonlinear medium is replaced by a single 3-level (artificial) atom, and the two input fields
are tuned to the $|g\ra-|e\ra$ and the $|e\ra-|s\ra$ transition frequencies. 
The experiment was performed with
coherent states, demonstrating an average cross-Kerr phase shift of $20^\circ$ per photon, surpassing the previous state-of-the-art of $0.01^\circ$ per photon for propagating light \cite{Matsuda09, Perrella13, Venkataraman13}. In parallel, a phase shift of $45^\circ$ was demonstrated with Rydberg atoms, as discussed in Sec.~\ref{sec-Rydberg}.

This work on the cross-Kerr effect has stimulated a great deal of theoretical
work analyzing the prospects of this system to be used for the QND detection of photons.
Interestingly, a first work concluded that QND detection could \textit{not} be
achieved with a single atom, due to atomic saturation effects
\cite{Fan13}. Because a bulk nonlinear crystal can largely be considered as an
incoherent ensemble of such atoms, this work strongly suggest that QND
detection is not possible in a bulk system, which was an important general
result. However, subsequent work has shown that careful arrangements of
multiple atoms, sequentially interacting with the propagating photon, can
achieve QND detection of a single propagating photon  \cite{Fan14, Sathyamoorthy14}.

The Chalmers group has also looked at a system with one atom in front of a
``mirror" \cite{Hoi14}.   In this case, the mirror-like boundary condition is
created by simply terminating the transmission line with a short circuit to
ground at one end.  The effective separation is modulated around the value of $\lambda/2$.  Interference between
the input field, the field scattered by the atom and the field reflected by
the mirror creates a standing wave pattern.  When the separation is exactly
$\lambda/2$, the atom sits at a node of the field, and it is effectively hidden
from the probe. This can be thought of as an interaction between the atom and
its image in the mirror, much in the same way that the two real emitters interact
in the Z{\"u}rich experiment \cite{Loo13} described below.  Interestingly, it was shown that the atom not only hides from the classical probe field, but also from vacuum fluctuations, with a suppression of the free-space relaxation rate by a factor of 50 being observed.  In contrast to the Purcell effect, where a cavity can be used to suppress the relaxation rate, the suppression achieved through this novel form of vacuum engineering occurs even though the atom is coupled to a continuum of propagating states.

While superconducting circuits interacting with microwave photons show great
promise for implementing quantum nodes, the implementation of long-distance quantum channels using optical photons is very far advanced, with records distances of $>100$ for free-space and in-fiber transmissions \cite{Ursin2007,Korzh2015}. This state of affairs has motivated considerable work in recent years towards the development of a quantum interface between microwave and optical photons. The fundamental
difficulty is the several orders of magnitude that separate the energy scales
of these two classes of photons.   A number of research groups have attempted
to bridge this gap in a wide variety of physical systems. One common approach
is to use ensembles of dopant atoms in a host crystal, taking advantage of the
fact that specific transition can be excited both through microwave and
optical Raman transitions. Ensembles are used to enhance the coupling to
single photons.  A number of different systems are being studied including
nitrogen-vacancy centers in diamond \cite{Kubo11, Julsgaard13, Grezes14,
  Amsuss11, Putz14} and rare-earth ions, especially erbium \cite{Bushev11, Staudt12, Afzelius13, Probst13}.  An entirely different approach that has shown promise is to use optomechanical systems, where mechanical vibrations (phonons) in nano- or micro-mechanical systems act as an intermediary between the microwave and optical photons \cite{Lin10, Andrews14, Bochmann13}. Finally, collective magnon excitations in macroscopic ferrimagnetic crystals are also being studied as a potential medium for an interface \cite{Tabuchi14}.  The shear range of physical systems being studied as a potential implementation of a quantum interface speaks to the compelling nature of the problem.

\section{Multiple emitters in the strong-coupling regime}
\label{MultEm}

\begin{figure}[tb]
\includegraphics[width=8.5cm]{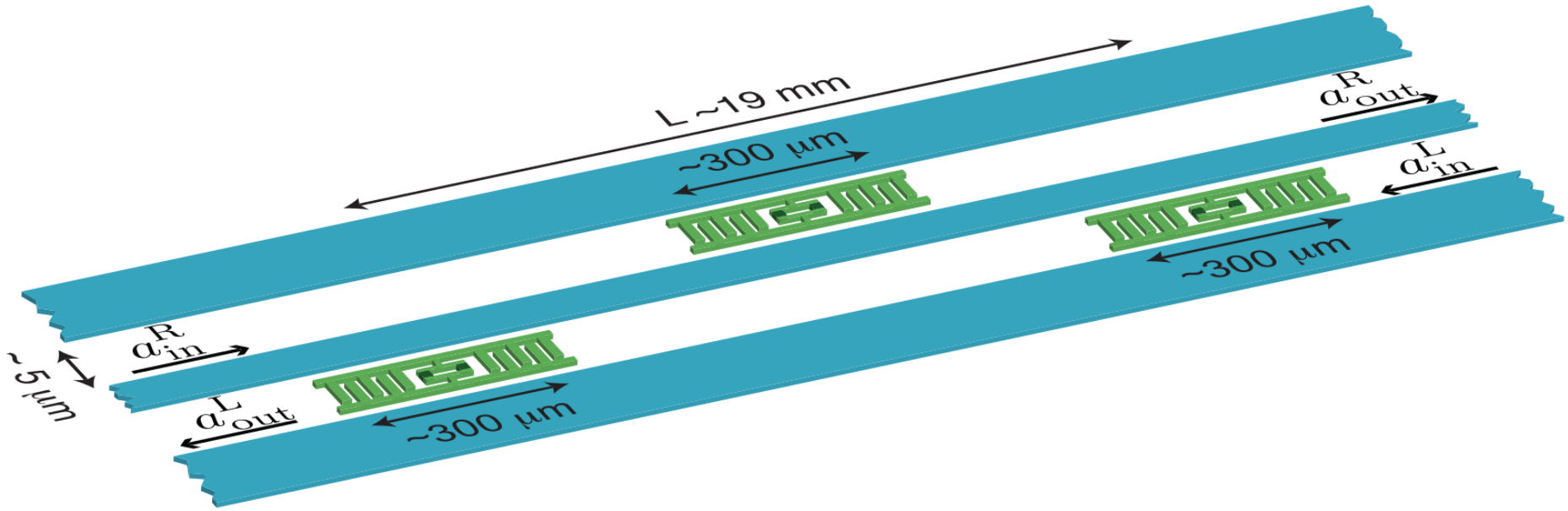}
\caption{(Color online). Photon-mediated interaction between distant emitters in
  superconducting circuits. Transmon qubits (length $\approx 300 \mu$m) acting as emitters are coupled to an open 1D transmission line. Adapted from \textcite{Lalumiere13}.}
\label{MultiEm}
\end{figure}

Investigating the interaction of propagating photons with multiple emitters in 1D continuum, as illustrated in Fig.~\ref{MultiEm}, is of fundamental and practical importance.
For example, photon-mediated interaction between distant emitters, as recently demonstrated in 1D \cite{Loo13,Lalumiere13}, can be used to generate long-range entanglement between the emitters \cite{Greenberg15, Gonzalez11, Zheng13a}.  More generally, we can imagine systems with multiple emitters being used in increasingly more advanced quantum communication nodes.
Moreover, strong coupling of multiple emitters with propagating photons confined below
the diffraction limit has potential practical applications for detection and
subwavelength imaging of atoms, ions, molecules, QDs, or color centers
in a host crystal \cite{Roy13a, Smolyaninov05,  Chen10, Kuhn06, Bushev11, Rezus12}.
Finally, systems of multiple emitters may realize \emph{photonic} simulators for the quantum many-body dynamics of electrons in condensed matter. Examples of analyzed phenomena in such simulators include a Luttinger liquid of photons \cite{Angelakis11}, quantum phase transitions of light \cite{Hartmann06, Greentree06}, optical Josephson interferometry \cite{Gerace09}, and the Tonks-Girardeau gas of
photons \cite{Chang08}.

The potential of multi-emitter systems has motivated a number of theoretical studies in recent
years. While most have been within the standard Markovian approximation
\cite{Yudson08, Roy13b, DzsotjanPRB10, ChangNJP12, Caneva15}, others have looked for distinctly non-Markovian effects \cite{Laakso14, Zheng13a}.

\subsection{Photon-mediated interaction between distant emitters}
The interaction of an isolated emitter with the vacuum fluctuations of the
electromagnetic field leads to
the spontaneous emission of real photons and to a renormalization of its
transition frequency (Lamb shift) via the emission and absorption of
virtual photons. A second emitter in the system can absorb both types of these photons,
giving rise to an effective interaction between the two emitters \cite{Goldstein97}. In a 3D space, the
strength of such interaction falls off rapidly with the separation between the emitters because of the
large mode volume of the photons and the resulting mode mismatch with the emitters. \cite{DeVoe96, Eschner01}.
As we saw for single emitters, these limitations are mitigated by confining 
the system to 1D.

Recently, the Z{\"u}rich group observed clear signatures of photon-mediated interaction between two superconducting transmon qubits 
\cite{Loo13}. The physical separation between the emitters
was large enough that there was no direct coupling between them. While the actual separation $d=18.6$~mm was fixed in the experiment, the normalized separation $d/\lambda$, measured in transition wavelengths, was changed by tuning the transition frequency of the emitter. The normalized separation could be tuned between $d/\lambda=1$ and $d/\lambda=3/4$.

\begin{figure}[tb]
\begin{tabular}{@{}cccc@{}}
\includegraphics[trim={0.17cm 3.93cm 0.13cm 0cm},clip,width=4.0cm]{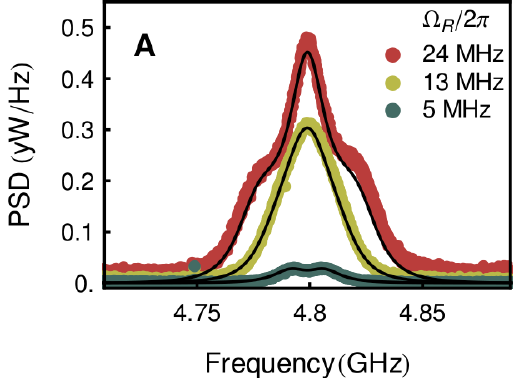}&&
\includegraphics[trim={0.17cm 0.05cm 0.13cm 3.9cm},clip,width=4.0cm]{Loo.pdf}
\end{tabular}
\caption{(Color online). Photon-mediated exchange interaction between two distant  superconducting qubits in an open transmission line.
(A and B) Power spectral density (PSD) of the resonance fluorescence of two qubits in resonance at $d \sim 3\lambda/4$, driven at the indicated Rabi rates,
$\Omega_R$. PSD falls with decreasing $\Omega_R$. At drive rates much lower than the relaxation rate, $\Omega_R/2\pi
\le$ 5 MHz, the observed double-peak structure reveals the effective
exchange interaction between the two qubits. Adapted from \textcite{Loo13}.}
\label{Exchange}
\end{figure}

 For  $d/\lambda = 1$, the two emitters are driven with the same amplitude and phase by any resonant field in the transmission line.  This led to the observation of a superradiant bright state and a subradiant dark state. For $d/\lambda\sim3/4$, one emitter is at a node of the propagating field when the other is at an antinode, suppressing these superradiant effects. However, an exchange interaction mediated by virtual photons is maximized. The anticrossing in the two qubit spectrum caused by this exchange interaction was observed in the resonance fluorescence spectrum of the driven two-emitter system, as shown in Fig.~\ref{Exchange}. \textcite{Lalumiere13} described the two-emitter system of the Z{\"u}rich group using a Markovian master equation along with
input-output theory \cite{Lehmberg70}, obtaining good agreement with the
experimental results.

We note that the coherent exchange mediated by photons can generate a high
degree of long-distance entanglement between the emitters \cite{Hensen15,Gonzalez11,
  DzsotjanPRB10,Fang14, Fang15} which is an important ingredient for quantum information science.

\subsection{Small ensemble of adjacent emitters}
Several studies have been carried out on small
ensembles of adjacent emitters with no intrinsic interaction.  (We consider intrinsic interactions in Sec. \ref{sec-Rydberg}.) \textcite{Rephaeli11} calculate
 2-photon scattering from a pair of adjacent 2LEs, predicting that the
fluorescence of the emitters is completely quenched for a proper choice of input
\cite{Zhou95}. \textcite{Roy13a} compared 1-photon and 2-photon scattering from two 2LEs versus a single V-type 3LE, showing that the two cases can be distinguished by the statistics of the scattered field. \textcite{Zapasskii13} successfully demonstrated this in optical spin-noise spectroscopy experiments.

\subsection{Directional photon propagation}

Strong transverse confinement of guided photons leads to
large intensity gradients on the wavelength scale. In this nonparaxial regime, spin (polarization) and orbital angular momentum of light are coupled, an effect known as spin-orbit coupling. In particular, the spin state varies within the transverse
plane and along the propagation direction in the waveguide. Interestingly, the local spin of
this strongly confined light can be orthogonal to the propagation direction.

Utilizing the spin-orbit coupling, several
experimental groups \cite{Petersen14,Le15} have recently demonstrated
nonreciprocal scattering of light from dipolar scatterers
(QDs, gold nanoparticles, alkali atoms) coupled to waveguides (photonic crystals, nanofibers).
The propagation direction of scattered light in such systems
depends on the spin direction of the incident light.
Control of the directionality of the scattering process with over $90\%$ efficiency has been achieved in such nanophotonic waveguide interfaces. \textcite{Mitsch14} have further shown directional
spontaneous emission of photons from emitters into a nanophotonic waveguide. \textcite{Bliokh15} argue that the transverse spin
in evanescent waves and the spin-controlled directional excitation of surface or
waveguide modes is analogous to the quantum spin Hall
effect.

Directional propagation due to collective scattering, distinct from the above effects, was theoretically studied by \textcite{Roy10a, Roy13b} in spatially asymmetric atom-waveguide interfaces. Two different structures were investigated using scattering theory: a 2LE directly coupled to two waveguides with different coupling strengths, and a chain of closely spaced 2LEs with varying transition frequencies \cite{FratiniPRL2014} along the chain. While the 1-photon transmission is reciprocal in these systems, the two or multiphoton transmission is nonreciprocal due to the broken spatial symmetry. These systems can thus act as optical diodes or isolators \cite{Jalas2013} for few photon states.

\section{Multiple interacting emitters in the weak-coupling regime: Rydberg polaritons}
\label{sec-Rydberg}
\begin{figure*}[tb]
\includegraphics[trim={0cm 0cm 1cm 0.5cm},clip,width=16cm]{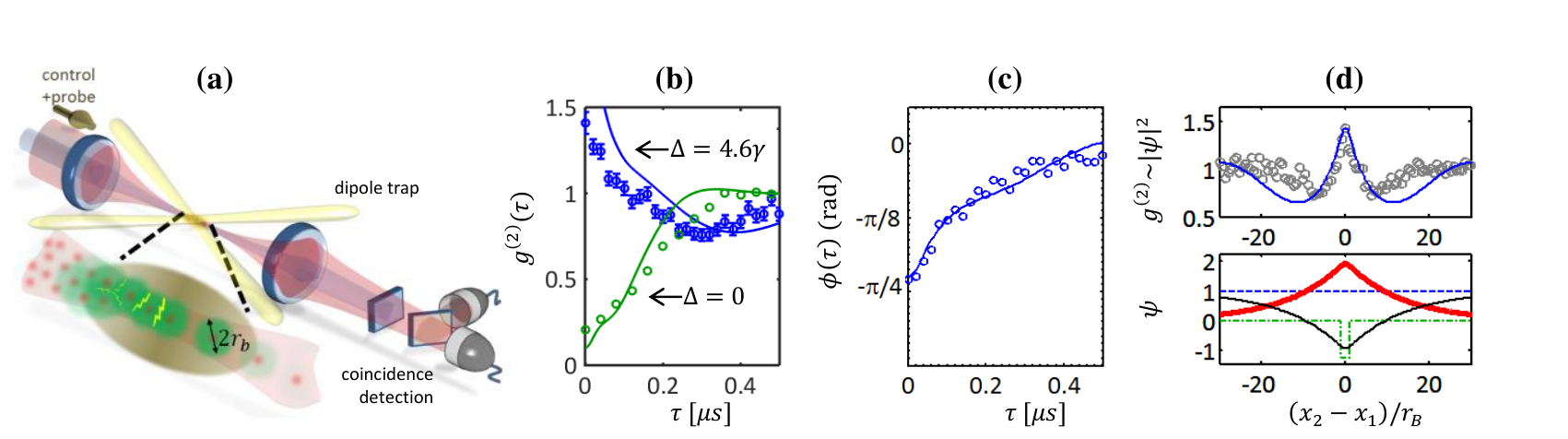}
\caption{(Color online). (a) Schematics of a typical Rydberg-polaritons setup: a cloud of ultracold alkali atoms is held between two confocal lenses. The outgoing probe photons are measured using single-photon detectors. (b) Normalized second-order correlation $g^{(2)}$ of the outgoing probe versus the time separation $\tau$ between the photons, showing the transition from antibunching in the dissipative regime ($\Delta=0$) to bunching in the dispersive regime ($\Delta=4.6\gamma$).
The group delay in the medium was 0.25 $\mu$s. Points are experimental data, lines are numerical simulations. (c) Conditional phase-shift obtained from a tomographic reconstruction of the outgoing 2-photon wavefunction.
(d) Signature of the 2-photon bound state: The interaction potential $U(r)$
is approximated by a well of width $2\rb$ (bottom, dash-dot line). The resulting bound state, observed in the shape of the measured $g^{(2)}$ (top, circles, where time is converted to distance via $x_2-x_1=\tau \veit$), conforms to that calculated using
Eq.~(\ref{eq_RydSch}) (top, solid line). The initial state $\psi(r,0)=1$ (bottom, dashed line) is a superposition of the bound state (bottom, thick line) and the manifold of scattering states (bottom, thin line). Adapted from \textcite{Peyronel12} and \textcite{Firstenberg13}.}
\label{fig_Rydberg}
\end{figure*}
Up until this point, we have considered explicitly only the interaction
between photons arising from their coupling to the same
emitter. In these configurations, efficient interaction requires the strong-coupling condition $\Gamma/\gamma \gg 1$, where the coupling $\Gamma$ of the emitter to the \emph{confined} probe channel is faster than all other relaxation rates $\gamma$, including spontaneous emission to the vacuum environment. In the alternative approach we shall now discuss, the interaction between photons is obtained without mode confinement in the opposite regime of weak coupling $\Gamma/\gamma \ll 1$, by utilizing many atoms, which in turn are intrinsically interacting.

In free space, the cumulative effect of $N$ atoms is often characterized by the resonant optical depth $\OD=2N\Gamma/(\gamma+\Gamma)=N\sigma_a/A$ \cite{Caneva15}. The atomic cross-section $\sigma_a$ is at most $\sim\lambda^2$ and typically much smaller than the focused beam area $A$, hence high $\OD$ are obtained only for large $N$. Taking $t_k$ from Eq.~(\ref{SBA11tk}), we recover a well-known expression for the total transmission amplitude $t_k^{N}=\exp(-\frac{\OD}{2}\frac{i\gamma}{\Delta+i\gamma})$ at $\Gamma/\gamma \ll 1$, where $\Delta$ is the detuning of the light from resonance. 
However contrary to the strong-coupling case, 
here most of the non-transmitted light  is lost rather than reflected. The ratio between the single-atom reflection coefficient $|r_k|^2$ from Eq.~(\ref{SBA11tk}) and the loss $1-|r_k|^2-|t_k|^2$ is $\Gamma/(2\gamma)\ll 1$.
Furthermore as we have seen, strong coupling is required for obtaining significant photon-photon correlation (cf. Fig.~\ref{2ndoc}). Indeed while high $\OD$ is useful for observing and utilizing single-photon effects, two more ingredients, namely the cooperative behavior of the atoms and suppression of loss, are required for strong photon-photon interactions in the weak-coupling regime.

Rydberg atoms provide such cooperativity via the dipole blockade mechanism \cite{LukinPRL2001}.
When exciting two or more atoms to Rydberg states --- electronic states with a large principal quantum number $n$ (typically $n=40 $ to $ 100$) --- the dipolar interaction between them shifts their Rydberg levels and consequently their excitation frequency \cite{RaithelPRA2007}.
Below a certain distance between the atoms, known as the blockade radius $\rb$ (typically $\le 10~\mu$m), the frequency shift is larger than the excitation linewidth (typically $1-10$ MHz for cold atoms), blocking the excitation of more than one atom. Such large $\rb$ are achieved at the van-der-Waals regime, where the frequency shift depends on the extremely large dipoles ($\sim 10^4$~Debye) to the power of 4.
The blockade effect itself can be used for various quantum information processes with atoms \cite{MolmerRMP2010, ZollerPRL2009}.

The consequence of the dipole blockade for propagating photons in quite intuitively understood. A single photon, exciting a single Rydberg atom, can block the excitation of all $N_\text{B}\gg 1$ atoms in the surrounding virtual sphere of radius $\rb$. If the optical depth of the blockade sphere $\ODB=N_\text{B}\sigma_a/A$ is large, it allows a single photon to significantly alter the optical response for other photons.

The final ingredient required for effective photon-photon interactions in the weak-coupling limit is the suppression of loss, which can be accomplished using EIT as proposed by \textcite{KurizkiPRA2005}.
A ladder-type configuration as in Fig.~\ref{sch}(b) is used, with a \emph{probe} photon exciting the ground-state atom to an intermediate state, which is coupled to the Rydberg state by a strong \emph{control} field with Rabi frequency $\Omega_c$. The probe is transmitted due to EIT within a narrow spectral window. By shifting this window, the dipole blockade disables the EIT for more than one probe photon in the blockade sphere. A strong dependence on the rate of incoming probe photons can be observed by monitoring the average transmission \cite{AdamsReview2013}. The optical nonlinear behavior and the emergence of non-classical correlations during propagation were theoretically investigated within the mean-field approximation by \textcite{AdamsPRL2010}, \textcite{PohlPRL2011}, \textcite{PetrosyanOtterbachPRL2011}, and others.

The transmitted probe experiences strong dispersion and thus reduced group velocity $\veit\ll c$, where $c$ is the speed of light in vacuum. It is therefore instructive to view each propagating photon as a slow light-matter polariton \cite{FleischhauerPRL2000}. Typically these polaritons have a negligible photonic component, of order $\veit/c$, guaranteeing the excitation of a Rydberg atom per each propagating photon at any given time. The strong dipolar forces between two Rydberg atoms thus effectively mediate interaction between the two propagating photons.

\subsection{Two-photon dynamics}

\textcite{Gorshkov11} provide a quantum description for two photons, which we now follow. A probe photon in the form of a polariton blocks EIT for the second probe photon, which experiences the response $t_k^{N_\text{B}}=\exp(-\frac{\ODB}{2}\frac{i\gamma}{\Delta+i\gamma})$ of bare two-level atoms. 
The nature of the photon-photon interaction is thus controlled by the detuning of the probe $\Delta$ from the intermediate atomic state.
On-resonance excitation ($\Delta=0$) leads to scattering of the second photon, inducing an effective \emph{dissipative} interaction between the photons. The second photon is scattered inside the blockade sphere with a probability $1-e^{-\ODB}$. 
In contrast, for off-resonance excitation ($|\Delta|\gg\gamma$), the second photon experiences much less scattering but acquires a nonzero phase $\phi=-(\gamma/\Delta)\ODB/2$. This leads to a conditional phase shift $\phi$ for two or more photons, inducing a \emph{dispersive} interaction between the photons. We see that for both on and off resonant processes, $\ODB$ is the key parameter determining the strength of the effective interaction.

Nonlinear optical response with Rydberg polaritons was demonstrated originally with $\ODB\ll 1$ by \textcite{AdamsPRL2010}. The limit of quantum nonlinearity was afterwards reached with $\ODB\approx 10$ by \textcite{Peyronel12} in a system illustrated in Fig.~\ref{fig_Rydberg}(a). The long axis of an elongated cloud of ultra-cold atoms was $L\approx100~\mu$m, and so naively could fit 10 blockade spheres of radius $\rb\approx 10~\mu$m, each containing $N_\text{B}\approx1000$~atoms. To render an effective 1D system, the light was focused to a diameter larger than the wavelength but smaller than $\rb$. In this way, paraxial diffraction did not substantially alter the transverse extent of the photon wavefunction along its propagation, and at the same time, two polaritons propagating side-by-side blocked each other.

Quantum tomography can be used to completely characterize the outgoing 2-photon state as a function of the time separation $\tau$ between the photons \cite{Firstenberg13}. In particular, it yields the second-order correlation function $g^{(2)}(\tau)$ 
and the conditional phase-shift $\phi(\tau)$ relative to the noninteracting case.

The dissipative interaction at $\Delta=0$ was observed by \textcite{Peyronel12} at an average level of less than one photon in the medium. The probability that two photons exit the medium simultaneously was suppressed by the blockade mechanism, resulting in photon antibunching $g^{(2)}(0)<1$, as seen in Fig.~\ref{fig_Rydberg}(b). The dispersive interaction at $\Delta\ne 0$ was demonstrated by \textcite{Firstenberg13}. In addition to a large conditional phase shift $\phi(0)=45^\circ$ [Fig.~\ref{fig_Rydberg}(c)], photon bunching $g^{(2)}(0)>0$ was observed [Fig.~\ref{fig_Rydberg}(b)].

The bunching in the dispersive regime can be attributed to an effective attractive force between the photons, intuitively arising from an increase in the group velocity inside the blockade volume. To better describe the 2-photon dynamics, one defines an effective 2-photon wavefunction inside a medium of length $L$ as a function of the coordinates $x_1$ and $x_2$ of the two photons
\bea
\psi(x_1,x_2)= \frac{\la \varphi| a(x_1)a(x_2) |\Psi \ra }{\la \varphi| a(x_1) |\Psi \ra \la \varphi| a(x_2) |\Psi \ra}.
\label{eq_psi_2p}
\eea
Here $|\Psi \ra$ is the full wavefunction of the system and $a(x)$ annihilates a photon at position $x$. We assume a stationary scenario with a constant incoming coherent state. Due to symmetry, it is useful to work with the relative $r=x_2-x_1$ and mean $R=(x_1+x_2)/2$ coordinates. From the definition (\ref{eq_psi_2p}), classical light in the absence of photon-photon interactions is described by $\psi(r,R)=1$; complete photon blockade corresponds to $\psi(0,L)=0$, which is measured in experiments using $|\psi(0,L)|^2=g^{(2)}(0)$; and a pure conditional phase $\phi$ corresponds to $\psi(0,L)=e^{i\phi}$.

\textcite{Firstenberg13} show for $\Omega_c\ll\Delta$ that $\psi(r,R)$ approximately follows a Schr\"odinger-like equation with $R$ playing the role of time,
\bea
i\frac{\partial \psi}{\partial R} =
\frac{4 l_\text{a} \Delta}{\gamma} \frac{\partial ^{2}\psi }{\partial r^{2}}+
\frac{\gamma}{ l_\text{a} \Delta} U(r)
\psi.\label{eq_RydSch}
\eea
Here $l_\text{a}=L/\OD=2\rb/\ODB$ is the attenuation length.
Assuming a repulsive van-der-Waals interaction, the effective potential can be approximated by the step function $U(|r|\le\rb)=1$.
The effective photon mass of this Schr\"odinger-like evolution originates from the quadratic component ($\propto k^2$) of the dispersion of individual polaritons. This component causes the change in group velocity when the Rydberg level shifts.

Equation (\ref{eq_RydSch}) 
approximately describes a potential well. Both the mass and the potential terms flip signs for $\Delta\rightarrow -\Delta$, so the effective force remains attractive. The 2-photon bound state of the finite-well potential, shown in Fig.~\ref{fig_Rydberg}(d), governs the evolution of an incoming wavefunction $\psi(r,0)=1$. In Fig.~\ref{fig_Rydberg}(d), in order to compare the measured $g^{(2)}(\tau)$ to the calculated $\psi(r,L)$, the time $\tau$ has been converted to distance $r$ using the group velocity $\veit=l_\text{a}\Omega_c^2/(2\gamma)$. In experiments, $\Omega_c^2\ll c\gamma/l_\text{a}$ and thus $\veit\ll c$, as we have assumed throughout this section.

Following this initial model, \textcite{Bienias14} have used scattering theory to study the 1D scattering properties for two photons for a wide range of the system parameters. By calculating the effective 1D scattering length, they predict the existence of scattering resonances analogous to Feshbach resonances in cold atoms, where the interaction turns from attractive to repulsive. 
For the experimental parameter regime described above, \textcite{Bienias14} generalize Eq.~(\ref{eq_RydSch}) to account for nonstationary (but slowly-varying) probe input, essentially replacing the term $\partial/\partial R$ by $\partial/\partial R+\partial/(\veit\partial t)$.

Lately, two more treatments for the system have been introduced. \textcite{Caneva15} model the atomic ensemble by a chain of 3LEs along a 1D waveguide, generalizing the input-output formalism presented in the previous sections. With this effective description, they recover the mean ensemble behavior (\emph{e.g.}, the optical depth and $\veit$) and provide a recipe for calculating the high-order correlations of the outgoing photons. In parallel, \textcite{Moos2015} write down an exact many-body model for the system, including the loss (scattering out of the system) and the paraxial propagation of the probe light, and provide the conditions under which the model is reduced to an effective many-body system of slow-light polaritons in 1D. They verify the model in the 2-photon case by comparing it to exact numerical simulations.

\subsection{Many photons}
Recently, two groups have demonstrated \emph{storage} of Rydberg polaritons for implementing a single-photon switch or transistor \cite{DurrPRL2014,HofferberthPRL2014,DurrPRL2014b}. In these experiments, a `gate' Rydberg polariton is first converted into a stationary excitation (a spin wave) of one Rydberg atom by turning off the control field. Subsequent `signal' polaritons, coupled to a different Rydberg state, are blockaded by the stored polariton. The number of signal photons gated by the stored excitation determines the `gain' of the transistor, recently reaching 100 \cite{HofferberthNatComm2016}.

So far, the many-body behavior of Rydberg polaritons has been limited to theoretical studies. \textcite{zeuthen2016correlated} describe the many-body evolution in the dissipative regime, where the system can act to transform a classical input to a regular train of single photons. \textcite{Bienias14} formulated a low-energy many-body Hamiltonian in the dispersive regime based on their derivation of the 1D scattering length. \textcite{OtterbachPRL2013} introduced an approximate Hamiltonian and used Luttinger liquid theory to predict the Wigner crystallization of Rydberg polaritons. To this end, they assumed an initial transient (preparation) phase where co-located polaritons are scattered out of the system, such that the photon-photon interaction is dominated by the tail of the repulsive van-der-Waals interaction.
\textcite{Moos2015} followed up on these arguments to derive their aforementioned effective many-body model and used numerical calculations (with the DMRG method) to compare with the Luttinger liquid result. 

\subsection{Limitations and prospects}

In a long medium ($L>2\rb$), a co-propagating pair of interacting photons traverses several blockade radii. The accumulated effect of the interaction can be calculated from Eq.~(\ref{eq_RydSch}) or from its generalization to the dissipative regime. Such calculations take into account the distortion (dispersion and absorption) of the photon wave-packet due to the finite bandwidth of the EIT transmission window, manifested in Eq.~(\ref{eq_RydSch}) by the mass term. The results depend on both $\OD$ and $\ODB$, with dissipative interaction ($\Delta=0$) yielding a blockade probability $1-\OD^{-1/2}e^{-\ODB}$ \cite{Peyronel12} and dispersive interaction ($\Delta\gg\gamma$) yielding a conditional phase shift $\phi\propto\sqrt{\OD}\times\ODB$
\cite{Firstenberg13}.

Nevertheless, while $\OD$ strengthen the overall effect as implied by the above expressions, it is now widely accepted that $\ODB\gg1$ alone is the key condition for high fidelity of quantum-information operations. Under this condition, the optimized arrangement is a medium completely residing within one blockade volume ($L\le 2\rb$) and thus $\OD=\ODB\gg1$. Intuitively in the co-propagating case, this arrangement guarantees that the wave-packets of both photons are fully contained within the interaction range ($\rb$), so that the fidelity is not reduced by partial entanglement of different parts of the wave-packets. In the case of the aforementioned photonic switch, the gating probability of a `signal' polariton by a stored `gate' polariton equals $1-e^{-\ODB}$ and the decoherence of the 'gate' during the operation scales as $e^{-\ODB}$ \cite{LesanovskyPRA2015,murray2016many}; the overall infidelity of the switch is thus governed by the inefficiency of storage and retrieval of the 'gate', scaling unfavorably as $1/\ODB=1/\OD$ \cite{GorshkovPRA2007}. For realizing strongly-correlated many-body states of co-propagating photons, such as the photon train or the Wigner crystal, the requirements on $\ODB$ become even more stringent \cite{OtterbachPRL2013,zeuthen2016correlated}.

So far, experiments have reached $\OD=\ODB=13$ \cite{Hofferberth_arxiv2016}. When using a single Rydberg level, \textcite{PfauNatComm2014} and \textcite{DurrPRL2014} predicted a limit $\ODB \lesssim 20$ due to formation of Rydberg molecules at high atomic densities.
A particularly promising approach to circumvent this limit is by tuning so-called F\"{o}rster resonances between two \emph{different} Rydberg levels \cite{HofferberthNatComm2016, Rempe2014b}.
Another approach is the introduction of an optical cavity with a finesse $\mathcal{F}$ around the atomic ensemble, where the condition $\ODB\gg 1$ is replaced by $\mathcal{F}\ODB \gg 1$ \cite{DasPRA2015}.
 Such enhancement of the coupling was first demonstrated by \textcite{Parigi12}, and recently long-lived cavity-Rydberg polaritons were realized \cite{SimonPRA2015}.

\section{Conclusions and perspectives}
\label{Conc}

Advances in both quantum optics and classical nonlinear optics were fueled by developments in laser, material, and electronic technologies, but apart from occasional intersections, these two fields remained almost separated for half a century. Research was mostly limited to the study of either high-intensity fields with macroscopic or mesoscopic materials or low-intensity fields with individual atoms. It is the recent development in realizing and understanding effective strong photon-photon interactions at low light intensities in various systems that triggered the new field of `quantum nonlinear optics' \cite{Chang14}.

There has been significant progress in theoretically describing nonlinear, out-of-equilibrium dynamics of scattered photons from individual emitters in a 1D continuum. A large set of all-optical quantum devices and logic, such as single-photon routers and switches, few-photon
optical diodes, and conditional phase gates, has been proposed using individual two- or multi-level
emitters in 1D. Some of these could find potential applications in future optical quantum circuits. However, our current understanding of light-matter interactions in the mesoscopic regime, with an ensemble of intrinsically interacting or
noninteracting emitters, and the ensuing strongly-correlated dynamics, is still at its infancy. This mesoscopic quantum regime was
addressed earlier in theoretical studies of quantum solitons in optical
fibers \cite{Drummond93}. Hydrodynamic, as well as microscopic descriptions,
were developed to investigate strong light-matter interaction in this
regime. However, the intrinsic nonequilibrium nature of the current experimental
systems calls for more careful investigation to obtain a fully quantum
mechanical theory in the mesoscopic regime. Numerical methods -- especially
time-dependent DMRG -- may play an important role.

 The dynamics of strongly confined propagating photons interacting with
 multiple emitters is similar to the nonequilibrium dynamics of many-body
 condensed matter systems, but still shows some important differences. One significant
 feature of photonic systems is the controllability over their parameters, including the
 strength of the interactions and dissipation. This makes the photonic systems suitable candidates for performing quantum
 simulation of condensed matter phenomena. Nevertheless, it is also of fundamental interest to investigate quantum nonlinear dynamics in
the regime where the number of particles (\emph{i.e.}, photons) is not conserved. This is a unique feature of photonic systems compared to natural closed material systems. In particular, it is interesting to
explore the role of loss and dissipation in the quantum-to-classical transition of mesoscopic many-body systems.
The interplay of spin-orbit interactions and the collective
scattering of light has not been studied so far, and it is an important direction
for future studies. In the presence of disorder,
these nonlinear systems have the potential to exhibit many-body localization of photons,
which has been very little explored theoretically or experimentally.

In a few short years, the experimental efforts with scattering of microwave photons from superconducting artificial atoms and with Rydberg polaritons in cold atomic ensembles have produced multiple interesting and important results. So far, the majority of the results with superconducting systems have involved only a single artificial atom. However these systems are very well suited for scaling up to more atoms. The standard nanofabrication techniques employed in making the devices studied here can easily produce devices of many atoms with well-controlled spacing, coupling, detunings, etc..  The range of possible experiments is diverse, ranging from exploring fundamental physics of mesoscopic atom-photon systems to more sophisticated devices for quantum nodes. In addition to increasing the system size, superconducting circuits have room to increase the atom-photon coupling strengths into new regimes that have not been explored experimental in any system.  In fact, ultrastrong coupling, where $\Gamma \sim \omega_e$, has recently been demonstrated experimentally \cite{Peropadre13, Sanchez14, Forn16}. The ultrastrong-coupling regime promises many new surprises.

At the same time, we saw that strong photon-photon interactions are also enabled in the weak-coupling regime by the action of atomic cooperativity in an ensemble. Rydberg atoms, in particular, exhibit such cooperativity via the blockade mechanism. The dipolar interaction between the Rydberg atoms conceptually transforms the blockade sphere to a two-level superatom, replacing the 2LE of the strong-coupling regime. In both regimes, the photon-photon interaction is governed by the 2-photon bound states and changes its nature depending on the frequency detuning. However, while a strongly-coupled 2LE preserves the photon number in the channel, the resonant response of the superatom is predominantly lossy  (while the superatom is transparent due to EIT for a single photons on resonance, it scatters the subsequent photons, which is exactly the opposite of a 2LE). Consequently, the reflection associated with a strongly-coupled 2LE is negligible in the superatom case. The large extent of the superatom relatively to the optical wavelength also contributes to the diminishing of reflection.

Some directions are being studied in both the 2LE and the Rydberg systems, such as nonlinear transmission at the few-photon level, switching and photonic transistors, conditional-phase gates, production of bunched or anti-bunched light, and deterministic entanglement of initially independent photons. 
The distinct features of the Rydberg systems make them suitable to pursue a variety of rich two-body phenomena, such as a finite-size photonic molecule and exotic many-body behavior, particularly photon crystallization. 
With the increasing understanding of the capabilities and the limitations of the experimental systems, further developments are soon to be implemented in the preparation of the atomic medium, the atomic excitation schemes, and the optical mode confinement. We expect the field to continue evolving alongside the single-emitter systems and predict that many of the existing ideas, and certainly more to come, will be realized in the near future.

\begin*{\it Acknowledgments.}
We are grateful to many colleagues for discussion and guidance, including C.~S.~Adams, H.~U.~Baranger, N.~Bondyopadhaya, H.~P.~B\"{u}chler, Y.-L.~L. Fang, A.~V.~Gorshkov, S.~Hofferberth, G.~Johansson, M.~D.~Lukin, F.~Nori, T.~Peyronel, T.~Pohl, V.~Vuletic, and S.~Xu. DR acknowledges funding from the Department of Science and Technology, India via the Ramanujan Fellowship. CMW acknowledges financial support from NSERC of Canada, Industry Canada and the Government of Ontario. OF acknowledges support by the Israel Science Foundation and ICORE, the European Research Council starting investigator grant Q-PHOTONICS 678674, the Minerva Foundation, the Sir Charles Clore research prize, and the Laboratory in Memory of Leon and Blacky Broder.
\end*

\appendix
\section{Derivation of the input-output formalism}\label{inoutform}
The effective Hamiltonian of a two-level emitter coupled to chiral
(right-moving) photon mode in 1D continuum is given by
\bea
H^c_{\rm io}/\hbar&=&\int_{-\infty}^{\infty} dk~\vg ka^{\dg}_{R,k}a_{R,k}+\f{1}{2}
\om_e \sigma_z\nn\\ &+&\frac{V}{\sqrt{2\pi}}\int_{-\infty}^{\infty} dk[a^{\dg}_{R,k}\sigma_-+\sigma_+a_{R,k}].
\eea
The Heisenberg equations of motion for the operators are
\bea
\f{da_{R,k}}{dt}&=&-i\vg ka_{R,k}-\frac{i V}{\sqrt{2\pi}}\sigma_-, \label{he1}\\
\f{d\sigma_-}{dt}&=&-i\omega_e\sigma_-+\frac{i V}{\sqrt{2\pi}}\int dk \sigma_z a_{R,k}, \label{he2} \\
\f{d\sigma_z}{dt}&=&\frac{2 i V}{\sqrt{2\pi}}\int dk (a^{\dg}_{R,k}\sigma_--\sigma_+a_{R,k}). \label{he3}
\eea
Equation (\ref{he1}) is formally solved by multiplying it by $e^{i\vg kt}$ and
integrating from an initial time $t_0<t$ to get
\bea
a_{R,k}(t)=a_{R,k}(t_0)e^{-i\vg k(t-t_0)}-\frac{i V}{\sqrt{2\pi}} \int_{t_0}^t dt'
\sigma_-(t')e^{i\vg k(t'-t)}.\nn\\\label{he4}
\eea
Next we integrate Eq.~(\ref{he4}) with respect to $k$ and introduce a field operator
\bea
\Phi(t)=b_{R, \rm in}(t)-i\f{V}{2\vg}\sigma_-(t),\label{he5}
\eea
where $\Phi(t)=(1/\sqrt{2\pi})\int dk a_{R,k}(t)$, and an input operator \cite{Fan10}
\bea
b_{R, \rm in}(t)=\f{1}{\sqrt{2\pi}}\int dk~a_{R,k}(t_0)e^{-i\vg k(t-t_0)}.
\eea
Plugging Eq.~(\ref{he5}) in Eqs.~(\ref{he2}) and (\ref{he3}), we get
\bea
\f{d\sigma_-}{dt}&=&iV\sigma_z b_{R, \rm
  in}(t)-\f{\Gamma}{2}\sigma_--i\omega_e\sigma_-,\label{he6}\\
\f{dN_e}{dt}&=&-iV(\sigma_+b_{R, \rm in}(t)-b^{\dg}_{R, \rm in}(t)\sigma_-)-\Gamma N_e,\label{he7}
\eea
where $\Gamma=V^2/\vg$ and $N_e=(\sigma_z+1)/2$. Similarly by integrating Eq.~(\ref{he1}) up to a final time $t_1>t$, we find
\bea
\Phi(t)=b_{R, \rm out}(t)+i\f{V}{2\vg}\sigma_-(t).\label{he8}
\eea
Finally we can write from Eqs.~(\ref{he5}) and (\ref{he8})
\bea
b_{R,\rm out}(t)=b_{R,\rm in}(t)-i\f{V}{\vg}\sigma_-(t).
\eea

\bibliographystyle{apsrmp4-1}
\bibliography{references5}
\end{document}